%% file: main.tex
\documentclass[12pt,preprint,preprint,nofootinbib]{revtex4}
\pdfoutput=1
\usepackage{epsf, array, color}
\usepackage{graphicx}
\usepackage{subfigure}
\usepackage{bbold}
\usepackage{amsmath}
\allowdisplaybreaks
\usepackage{amssymb}
\usepackage{mathtools}
\usepackage{epigraph}
\usepackage[utf8]{inputenc}
\usepackage{graphicx}
\usepackage{textcomp}
\usepackage{lscape}
\usepackage{pdflscape}
\usepackage[linktocpage,colorlinks,urlcolor=blue,hyperfootnotes=true]{hyperref}

\usepackage{url}

\usepackage{dcolumn}

\def\no{\nonumber\\}
\def\fn{\footnote}
\def\be{\begin{equation}}
\def\ee{\end{equation}}
\def\ba{\begin{alignedat}}
\def\ea{\end{alignedat}}
\def\bea{\begin{eqnarray}}
\def\eea{\end{eqnarray}}
\newcommand{\bs}{\begin{subequations}}
\newcommand{\es}{\end{subequations}}

\newcommand{\mF}{\mathcal{F}}
\newcommand{\mL}{\mathcal{L}}
\newcommand{\mO}{\mathcal{O}}

\newcommand{\cba}{c_{\beta-\alpha}}

\usepackage{color} 
\definecolor{mygreen}{RGB}{0,128,0}

\definecolor{raq}{RGB}{255,164,0}

\usepackage{ulem}
\usepackage{soul}

\begin{document}

\title{Matching the 2HDM to the HEFT and the SMEFT: \\ Decoupling and Perturbativity}

\renewcommand*{\thefootnote}{\fnsymbol{footnote}}
\author{Sally Dawson$^{a}$\footnote{dawson@bnl.gov}}
\author{Duarte Fontes$^{a}$\footnote{dfontes@bnl.gov}}
\author{Carlos Quezada-Calonge$^b$\footnote{cquezada@ucm.es}}
\author{Juan José Sanz-Cillero$^b$\footnote{jjsanzcillero@ucm.es}}
\affiliation{
\vspace*{.5cm}
  \mbox{$^a$Department of Physics,\\
  Brookhaven National Laboratory, Upton, N.Y., 11973,  U.S.A.}  
  \mbox{$^b$Departamento de Física Teórica and IPARCOS,
Universidad Complutense de Madrid,}
\mbox{Plaza de las Ciencias 1, 28040-Madrid, Spain}
\vspace*{.5cm}
  }

\begin{abstract}
We consider the 2 Higgs Doublet Model (2HDM) and compare two effective field theory (EFT) approaches to it, according to whether the heavy degrees of freedom are integrated out before (SMEFT) or after (HEFT) spontaneous symmetry breaking.
By requiring decoupling and perturbativity in the 2HDM, we define a consistent EFT expansion in inverse powers of the heavy masses which is 
applied to
both the SMEFT and the HEFT tree level matchings to the 2HDM. We organize this  expansion with a dimensionless parameter $\xi$, and investigate the tree-level scatterings $hh\rightarrow hh$ and $WW\rightarrow hh$ up to $\mathcal{O}(\xi^2)$. We find no differences between the HEFT and the SMEFT approaches at this order.
We show scenarios where even 
including dimension-8 operators of the SMEFT is insufficient to obtain an accurate matching to the 2HDM.

\end{abstract}

\preprint{IPARCOS-UCM-23-034}

\maketitle

\renewcommand*{\thefootnote}{\arabic{footnote}}
\setcounter{footnote}{0}

\section{Introduction}
\input{intro.tex}

\section{2HDM}
\input{2hdm}

\section{Decoupling and perturbativity}
\input{perturbativity}

\section{SMEFT}
\input{smeft}

\section{HEFT}
\input{heft_new}

\section{Results}
\input{results}

\section{Conclusions}
\input{conclusions}

\section*{Acknowledgments}
We thank Howard Haber for discussions.
D.F. is also grateful to Ilaria Brivio, Matthew Sullivan and Robert Szafron for discussions. C.Q.C is grateful to Gerhard Buchalla for discussions about the singlet extension and Antonio Dobado on decoupling. 
S. D. and D. F. are supported by the U.S. Department of Energy under Grant Contract No. DE-SC0012704. C.Q.C has been funded by the MINECO (Spain) predoctoral grant
BES-2017-082408.   
This work was supported in part by Spanish MICINN No. PID2019–108655 GB-I00 Grant, Universidad Complutense de Madrid under research group 910309 and the IPARCOS institute.
Digital data pertaining to the HEFT matching is contained in the auxilliary file submitted with this paper.

\appendix 

\section{Further details on the 2HDM}
\input{appendix-2HDM}

\section{A note on the $\mathbb{Z}_2$ symmetric singlet extension of the SM}

\input{appendix-singlet}

\bibliographystyle{utphys}
\bibliography{refs}
\end{document}

%% file: intro.tex
\label{introduction}

Since the discovery of the Higgs boson in 2012, the experimental focus in the electroweak sector at the LHC has turned to precision measurements of Higgs observables and the search for heavy Higgs-like particles. To date, no significant deviation from the Standard Model (SM) predictions has been observed, suggesting that beyond the SM (BSM) physics, if it exists, must be at a much higher energy scale than that probed at the LHC. In this scenario, effective field theories (EFTs) are the tool of choice in the search for deviations from the SM. In principle, the EFTs represent a model-independent formalism which can then be matched to the predictions of specific ultraviolet (UV) complete models. 

Two types of EFTs can be used to model the unknown BSM physics that potentially affects the Higgs sector; they are the SM Effective Field Theory (SMEFT)~\cite{Weinberg:1979sa,Buchmuller:1985jz,Leung:1984ni} and the Higgs Effective Field Theory (HEFT)~\cite{Feruglio:1992wf,Bagger:1993zf,Koulovassilopoulos:1993pw} (cf. ref.~\cite{Brivio:2017vri} for a review). Both use exclusively SM degrees of freedom, and both
are invariant under the SM gauge groups
$\rm SU(3)\times SU(2)_L\times U_Y(1)$. However, while the SMEFT considers the Higgs field $h$ and the electroweak (EW) would-be Goldstone bosons, $\omega^a$, to be embedded in the $\rm SU(2)_L$ Higgs doublet, the HEFT treats $h$  as a gauge singlet and classifies the $\omega^a$ as an $\rm SU(2)_L$ triplet.
As a consequence, the SMEFT starts from the SM as it is before spontaneous symmetry breaking (SSB) of $\rm SU(2)_L\times U_Y(1) \to U_{EM}(1)$, and adds to it a tower of higher dimensional operators, $O_{i}^{n}$, built out of the (before-SSB) SM fields:
\be
\mathcal{L}_{\rm SMEFT} = \mathcal{L}_{\rm SM} + \sum_{n,i} \dfrac{C_{i}^{n} O_{i}^{n}}{\Lambda^{n-4}},
\ee
where $n>4$ is the 
dimension of the operator, $C_{i}^{n}$ are coefficients (usually known as Wilson coefficients, WCs) and $\Lambda$ the UV scale. 
By contrast, the HEFT starts by treating $h$ and the $\omega^a$ separately, in such a way that the latter are embedded into a unitary matrix $U$.
Moreover, the HEFT is an expansion in the number of covariant derivatives; 
at the lowest order, the part of the HEFT Lagrangian relevant for the
scattering processes discussed in this article is%
\fn{Only terms relevant for our current purposes are shown. In particular, fermions will not be relevant, and will be omitted in what follows.}
\be
\mL_{\rm HEFT} \supset \frac{v^2}{4} \mF(h) {\rm Tr}\left\{D_\mu U^\dagger D_\mu U\right\} +\frac{1}{2}(\partial_\mu h)^2 - V(h)   
,
\label{eq:heftdef}
\ee
where $v=246$ GeV represents the vacuum expectation value (vev) of the Higgs field
in the SM, $D_\mu$ is 
the covariant derivative, and
$\mF(h)$ and $V(h)$ are generic functions of $h$.
In general, one has $D_\mu U= \partial_\mu U + i g W_\mu^a\frac{\sigma^a}{2}U-i g' U \frac{\sigma^3}{2}B_\mu $, with $U=1$ in the unitary gauge.
The fact that  $h$ is a gauge singlet means that 
symmetry invariance allows $\mF(h)$ and $V(h)$ to be
to be   arbitrary  power series in $h$. 
Considering again the lowest order HEFT Lagrangian,
 we find
\be
\mathcal{F}(h) = 1 + 2 a \dfrac{h}{v} + b \dfrac{h^2}{v^2} + ...\, ,
\qquad 
V(h) = \frac{1}{2} m_h^2 h^2 \left( 1 +d_3 \frac{h}{v} +\frac{d_4}{4}\frac{h^2}{v^2} \, +... \right)\, , 
\label{eq:flare}
\ee
where $m_h$ is the $h$ mass, the dots stand for terms with higher powers of $h$,
and $a, b, d_3$ and $d_4$ are arbitrary couplings. These are normalized so that the SM is recovered when both $a=b=d_3=d_4=1$ and the remaining terms with higher powers of $h$ are set to zero.

A significant effort has been made in recent years to derive techniques to distinguish the SMEFT and the HEFT from one another from a pure bottom-up approach, i.e. without assuming knowledge about any possible BSM model~\cite{Alonso:2015fsp,Alonso:2016oah,Falkowski:2019tft,Gomez-Ambrosio:2022qsi,Gomez-Ambrosio:2022why,Cohen:2021ucp,Cohen:2020xca}.
Yet, since the EFTs are ultimately effective descriptions of a particular UV model, it is also relevant to discuss a top-down approach. In this case, the BSM model is assumed to be known, and a matching between  the EFTs and the UV model is obtained by integrating out the heavy degrees of freedom.
This exercise has been done in the recent literature especially for the SMEFT, considering several different UV models~\cite{Perez:1995dc,
Englert:2014uua,
Brehmer:2015rna,
Gorbahn:2015gxa,
Buchalla:2016bse,
Belusca-Maito:2016dqe,
Dawson:2017vgm,
Corbett:2017ieo,
Dawson:2020oco,
Jiang:2018pbd,
Haisch:2020ahr,
Dawson:2021jcl,
Dawson:2021xei,
Anisha:2021hgc,
Dawson:2022cmu}. 

In this paper, we follow the top-down approach taking the 2 Higgs Doublet Model (2HDM)~\cite{Lee:1973iz} as the BSM model, and discuss the matching to both the SMEFT and the HEFT. Ref.~\cite{Buchalla:2016bse} performed an exercise along these lines, choosing as the BSM model a singlet extension of the SM with a $\mathbb{Z}_2$ symmetry. It turns out that this model is very special, as it allows an EFT expansion which is exclusively governed by inverse powers of the heavy mass. By contrast, and as we will show, a consistent EFT approach cannot be applied to a model like the 2HDM unless one makes further assumptions besides those related to the physical masses. This aspect is intimately related to the notions of decoupling and perturbativity, which shall be discussed in detail below.

We will focus on the tree-level scattering processes $WW\rightarrow hh$ and $hh\rightarrow hh$, where the HEFT and SMEFT may have potential differences when matched to the 2HDM.
We 
pay particular attention to performing consistent expansions in the different EFTs, and investigate
how accurately they reproduce
the results of the 2HDM.

This paper is organized as follows. We start by recapping the 2HDM in section~\ref{2hdm}. Section~\ref{perturbativity} is  devoted to the notion of decoupling and to its consequences for an EFT expansion.
That allows us to study the SMEFT and HEFT matchings to the 2HDM, which we do in sections~\ref{smeft} and~\ref{heft}, respectively. Finally, we present our results in section~\ref{results} and our conclusions in section~\ref{conclusions}.
We provide further details on the 2HDM and on the model of ref.~\cite{Buchalla:2016bse} in the appendices.

%% file: 2hdm.tex
\label{2hdm}

For this review
of the 2HDM, we follow Ref.~\cite{Dawson:2022cmu} closely  (for more details, cf. Refs.~\cite{Gunion:1989we,Branco:2011iw}). The model adds an extra doublet $\Phi_2$ to the SM scalar doublet $\Phi_1$, and we define their vevs as $v_2/\sqrt{2}$ and $v_1/\sqrt{2}$, respectively (we take them to be real).
We impose a softly broken $\mathbb{Z}_2$ symmetry, under which the scalar doublets transform as $\Phi_1 \to \Phi_1$ and $\Phi_2 \to -\Phi_2$, whereas the fermion fields can transform in four different ways (each one corresponds to a different type of 2HDM).%
\footnote{Since the fermions will not be the focus of this paper, we refer the reader to ref.~\cite{Branco:2011iw} for details on the different types of 2HDMs.}
It is convenient to introduce an angle $\beta$ such that $t_\beta = v_2/v_1$, which allows us to move to the Higgs basis~\cite{Donoghue:1978cj,Georgi:1978ri,Botella:1994cs,Branco:1999fs} as:%
\fn{Here and in the following, it should be clear that, for any angle $x$, we use $c_x \equiv \cos(x), ~s_x \equiv \sin(x), ~t_x \equiv \tan(x)$.}
\be
\label{eq:basis-rot}
\left(\begin{array}{c}
H_{1} \\
H_{2}
\end{array}\right)=\left(\begin{array}{cc}
c_{\beta} & s_{\beta} \\
-s_{\beta} & c_{\beta}
\end{array}\right)\left(\begin{array}{c}
\Phi_{1} \\
\Phi_{2}
\end{array}\right).
\ee
In the Higgs basis, the second doublet ($H_2$) has no vev, whereas $H_1$ has the vev $v/\sqrt{2}$, with $v \equiv \sqrt{v_1^2 + v_2^2} = 246$ GeV.
Among the terms of the Lagrangian, we focus on just two, $\mathcal{L}_{\mathrm{2HDM}} \ni \mathcal{L}_{\mathrm{kin}} -V$, 
the former being the scalar kinetic piece and the latter the potential. In the Higgs basis, they read:
\bs
\bea
\label{eq:kinetic}
\mathcal{L}_{\mathrm{kin}} &=& \left(D_{\mu} H_1\right)^{\dagger} \left(D^{\mu} H_1\right) + \left(D_{\mu} H_2\right)^{\dagger} \left(D^{\mu} H_2\right),\\[3mm]
\label{eq:potential}
V &=& Y_1 H_{1}^{\dagger} H_{1}
+ Y_2 H_{2}^{\dagger} H_{2}+\left(Y_3 H_{1}^{\dagger} H_{2}+\textrm{h.c.}\right) \no
&&+ \frac{Z_{1}}{2}\left(H_{1}^{\dagger} H_{1}\right)^{2}+\frac{Z_{2}}{2}\left(H_{2}^{\dagger} H_{2}\right)^{2}+Z_{3}\left(H_{1}^{\dagger} H_{1}\right)\left(H_{2}^{\dagger} H_{2}\right)+Z_{4}\left(H_{1}^{\dagger} H_{2}\right)\left(H_{2}^{\dagger} H_{1}\right) \no
&& + \left\{\frac{Z_{5}}{2}\left(H_{1}^{\dagger} H_{2}\right)^{2}+Z_{6}\left(H_{1}^{\dagger} H_{1}\right)\left(H_{1}^{\dagger} H_{2}\right)+Z_{7}\left(H_{2}^{\dagger} H_{2}\right)\left(H_{1}^{\dagger} H_{2}\right)+ \textrm{h.c.}\right\},
\eea
\es
in such a way that, on the one hand, the minimization equations imply:
\be
\label{eq:theYs}
Y_1 = - \dfrac{Z_1}{2} v^2,
\qquad
Y_3 = - \dfrac{Z_6}{2} v^2 \,,
\ee
and, on the other, the $\mathbb{Z}_2$ symmetry (which is only explicit in the  basis of $\Phi_1$, $\Phi_2$) is manifested by the circumstance that only 5 of the 7 $Z_i$ are independent.
Although the parameters $Y_3, Z_{5}, Z_{6}, Z_{7}$ are in general complex (the remaining parameters are real by hermiticity), we restrict ourselves to the 
solution in which they have real values.%
\footnote{As stressed in Ref.~\cite{Fontes:2021znm}, though, one should keep in mind that those parameters are generally complex, since issues with renormalization would otherwise follow.}
CP symmetry is thus preserved at the leading order in the scalar sector, in which case 
$H_1$ and $H_2$ can be parameterized as:
\begin{align}
\label{eq:Higgs_basis_param}
H_1 = 
\begin{pmatrix}
G^+ \\
\frac{1}{\sqrt{2}}(v + h_1^{\mathrm{H}} + i G_0)
\end{pmatrix},
\hspace{3mm}
H_2 = 
\begin{pmatrix}
H^+ \\
\frac{1}{\sqrt{2}}(h_2^{\mathrm{H}} + i A)
\end{pmatrix},
\end{align}
with $h_1^{\mathrm{H}}, h_2^{\mathrm{H}}, G_0$ and $A$ real fields, and $G^+, H^+$ complex ones. With the exception of $h_1^{\mathrm{H}}, h_2^{\mathrm{H}}$, all of these states are already mass states ($G_0$ and $G^+$ are the would-be Goldstone bosons, and $A$ and $H^+$ are the pseudo-scalar and the charged scalar bosons, respectively). The mass matrix for $h_1^{\mathrm{H}}$ and $h_2^{\mathrm{H}}$ can be diagonalized by introducing a mixing angle $\alpha$ such that:
\be
\label{eq:diagonalization}
\left(\begin{array}{c}
h \\
H
\end{array}\right)
=
\left(\begin{array}{cc}
s_{\beta-\alpha} & c_{\beta-\alpha}\\
c_{\beta-\alpha} & - s_{\beta-\alpha}
\end{array}\right)
\left(\begin{array}{c}
h_1^{\mathrm{H}}\\
h_2^{\mathrm{H}}
\end{array}\right),
\ee
where $h$ and $H$ are the neutral scalar mass states, with $h$ being the scalar that is observed at the LHC.
Finally, defining the masses of $h, H, A$ and $H^{\pm}$ to be $m_{h}, m_{H}, m_A$ and $m_{H^{\pm}}$, respectively, we shall take the following parameters as independent:
\be
\label{eq:indep-real}
c_{\beta \! - \! \alpha},
\,
\beta,
\,
v,
\,
m_{h},
\,
Y_2,
\,
m_{H},
\,
m_A,
\,
m_{H^{\pm}}.
\ee
The expressions for the $Z_i$ parameters in terms of the independent parameters can be found in Appendix~\ref{2hdm-appendix}.

%% file: perturbativity.tex
\label{perturbativity}

In the following sections, we shall derive EFTs for the model described in section~\ref{2hdm}, which is taken as our UV model. Such a derivation requires a separation of scales in the UV model.
Let us  focus on the UV model, and assume that it has two disparate mass scales, $\Lambda$ and $v$, such that $\Lambda \gg v$. Intuition leads to the expectation that the physical effects of the particle(s) with mass of $\mathcal{O}(\Lambda)$ should be suppressed at low energies, i.e. at $\mathcal{O}(v)$. This is, in fact, the main idea of \textit{decoupling}, which is formalized in the Appelquist-Carazzone decoupling theorem~\cite{Appelquist:1974tg} (see also ref.~\cite{Brivio:2017vri}).

Yet there is an important caveat here. 
The decoupling theorem was formulated in ref.~\cite{Appelquist:1974tg} for a model without SSB, where the masses are independent parameters in the Lagrangian; in particular, they are independent of  interaction couplings. It follows that a given mass can be rendered very large (of $\mathcal{O}(\Lambda)$) without affecting the interaction couplings --- and, in particular, without requiring these couplings to become very large. In this way, taking a particle to be very heavy in a model without SSB does not jeopardize \textit{perturbativity}, which is an implicit assumption of the decoupling theorem.%
\fn{In this paper, we assume that decoupling requires perturbativity, and we do not consider the scenario in which the UV model violates perturbativity.}

In models with SSB, the situation changes considerably~\cite{Toussaint:1978zm,Veltman:1977kh,Collins:1978wz,Einhorn:1981cy,Haber:1993wf,Dobado:2002jz}. The reason is that particles in models with SSB often get their masses from the product of a (fixed) vev and an interaction coupling. To 
obtain a very heavy mass for a particle, one would thus need to take the interaction coupling to be very large --- which would, however, inevitably make perturbation theory invalid. Therefore, decoupling is not possible in this scenario: one cannot take a particle to be 
infinitely massive without violating perturbativity (see also the discussion in ref.~\cite{Brivio:2017vri}). 

It should be clear, on the other hand, that this does not mean that decoupling is impossible in a model with SSB. For it may happen that, in such theory, a 
particle gets at least \textit{part} of its mass from a mass parameter of the Lagrangian --- which, as mentioned above,
is independent of the remaining Lagrangian parameters, and in particular of the interaction ones. Hence, by taking that mass parameter to be very heavy (while keeping the interaction parameters fixed), one renders the particle at stake to be very massive, without endangering the validity of a perturbative description.   

We can apply this discussion to the 2HDM described in the previous section, which is the focus of this paper. Our goal is to make the particles which do not belong to the SM ($H$, $A$ and $H^\pm$) very heavy, so that an EFT for the 2HDM can be build using solely the degrees of freedom of the SM.
To that end, we must have:
\be
\label{eq:decoupling_rough}
m_{H} \simeq m_{A} \simeq m_{H^{+}}
\gg m_{h} = 125 \, \, {\rm GeV}.
\ee
To see how this can be obtained in a consistent way, it is convenient to write these masses in terms of $v$, $c_{\beta-\alpha}$ and parameters of the potential:
\bs
\label{eq:2HDM-masses-sq}
\bea
\label{eq:2HDM-masses-sq-a}
m_{h}^2 &=& \dfrac{c_{\beta-\alpha}^2}{2 \, c_{\beta-\alpha}^2 - 1} \, Y_2 + 
 \dfrac{2 \, (c_{\beta-\alpha}^2 -1) Z_1 + c_{\beta-\alpha}^2 Z_{345}}{4 \, c_{\beta-\alpha}^2 - 2} \, v^2 , \\
\label{eq:2HDM-masses-sq-b}
m_{H}^2 &=& \dfrac{(c_{\beta-\alpha}^2-1)}{2 \, c_{\beta-\alpha}^2 - 1} \, Y_2 + 
 \dfrac{c_{\beta-\alpha}^2 (2 Z_1 + Z_{345}) - Z_{345}}{4 \, c_{\beta-\alpha}^2 - 2} \, v^2, \\
\label{eq:2HDM-masses-sq-c}
m_{A}^2 &=& Y_2 + \dfrac{Z_{345} - 2 Z_5}{2} \, v^2 , \\
\label{eq:2HDM-masses-sq-d}
m_{H^{+}}^2 &=& Y_2 + \dfrac{Z_3}{2} \, v^2\, ,
\eea
\es
with $Z_{345} \equiv Z_3 + Z_4 + Z_5$.
As suggested above, each of the squared masses ($m_{h}^2$ included) contains two parts: one of them proportional to a mass parameter of the Lagrangian ($Y_2$), the other one proportional to the product between interaction couplings ($Z_i$) and the squared vev ($v^2$).
This means that $Y_2$ plays a fundamental role in decoupling, as it can be used to render $m_{H}$, $m_{A}$ and $m_{H^{+}}$ very large without compromising the validity of the perturbation theory.%
\fn{This also shows that, in a 2HDM with an exact (i.e. not softly broken) $\mathbb{Z}_2$ symmetry, it is not possible to decouple $H$, $A$ and $H^+$. The reason is that, in that case, $Y_2 \sim \mathcal{O}(Z_i v^2)$, so that $Y_2$ cannot be taken to be very large without violating perturbativity~\cite{Asner:2013psa}.}
It is also clear that, if eq.~(\ref{eq:decoupling_rough}) is to be obeyed, and if $c_{\beta-\alpha}$ is chosen as an independent parameter, then taking $Y_2$ to be very large is not enough; more than that, $c_{\beta-\alpha}$ must behave so as to ensure that $m_{h}$ stays fixed as $Y_2$ is increased. Another way to realize this
is to consider eqs.~(\ref{eq:theZs-real}), which show the $Z_i$ parameters written in terms of the parameters of eq.~(\ref{eq:indep-real}). From those equations (in particular eq.~(\ref{eq:Z1})), it is clear that the only way eq.~(\ref{eq:decoupling_rough}) can hold without having 
large $Z_i$ (i.e. without violating perturbativity) is to require that $c_{\beta-\alpha}$ scales with $\Lambda^{-2}$.

All of this leads us to define the \textit{decoupling limit} of the 2HDM~\cite{Haber:1989xc,Gunion:2002zf,Haber:2006ue,Asner:2013psa} --- which ensures eq.~(\ref{eq:decoupling_rough}) while complying with $Z_i/(4\pi) \lesssim \mathcal{O}(1)$ --- as:%
\fn{Although $Y_2$ could be written $Y_2 = \Lambda^2 + \Delta Y_2$, with $\Delta Y_2$ a real parameter, the latter can be set to zero without loss of generality.}
\bs
\label{eq:decoupling_def}
\bea
\label{eq:decoupling_def_a}
&Y_2 = \Lambda^2,
\qquad
m_{H}^2 = \Lambda^2 + \Delta m_{H}^2,
\quad
m_{A}^2 = \Lambda^2 + \Delta m_{A}^2,
\quad
m_{H^{+}}^2 = \Lambda^2 + \Delta m_{H^{+}}^2,& \\
&
\label{eq:decoupling_def_b}
\Lambda^2 \gg v^2, \qquad
m_{h}^2 \sim \mathcal{O}(v^2),
\qquad
\Delta m_{H}^2, \Delta m_{A}^2, \Delta m_{H^{+}}^2 \sim \mathcal{O} (v^2),& \\
\label{eq:scaling_of_camb}
&c_{\beta-\alpha} \sim \mathcal{O} (v^2/\Lambda^2),& 
\eea
\es
with $\Delta m_{H}^2$, $\Delta m_{A}^2$ and $\Delta m_{H^{+}}^2$ real parameters. 
In the following sections, the decoupling limit defined in eqs.~(\ref{eq:decoupling_def}) will be used to build \textit{expansions},
corresponding to either the SMEFT or the HEFT. This can be more easily done if we introduce an auxiliary dimensionless parameter $\xi$, which acts as the \textit{de facto} expansion parameter.
Then, assuming eq.~(\ref{eq:decoupling_def_a}), 
we implement the scaling  $v^2/\Lambda^2,\, \cba\, \sim {\cal{O}}(\xi)$ of eqs.~(\ref{eq:decoupling_def_b}) and~(\ref{eq:scaling_of_camb})  at the practical level
(in our codes)
through: 
\be
\label{eq:scaling}
\dfrac{1}{\Lambda^2} \to \dfrac{\xi}{\Lambda^2}\, ,
\qquad
c_{\beta-\alpha} \to \xi \, c_{\beta-\alpha}\, ,
\ee
while all the other scales and parameters are $\mO(\xi^0)$ and are left untouched.%
\fn{Ref.~\cite{Dittmaier:2021fls} followed a similar procedure. One might wonder whether it would be possible to have alternatives to eq.~(\ref{eq:scaling}) for which $\beta$ might also scale in a non-trivial way. Yet, by considering eqs.~(\ref{eq:Z2}) and~(\ref{eq:Z7}), it is easy to conclude that any scaling of $\beta$ would violate perturbativity.}
In this way, the expansion will correspond to a series of
positive powers of $\xi$. 
The trivial order --- $\mathcal{O}(\xi^0)$ --- implies the alignment limit, which is defined by $c_{\beta-\alpha} \to 0$ and corresponds to the scenario in which the $h$ couplings are exactly those of
the SM. In this way, decoupling implies alignment.%
\fn{The reverse is in general not true: it is possible to have alignment without decoupling~\cite{Gunion:2002zf,Carena:2013ooa,Bernon:2015qea}.
An EFT approach to the 2HDM  in general does not work in this case~\cite{Belusca-Maito:2016dqe}.}

Several aspects are worth mentioning here. The first one is that eq.~(\ref{eq:decoupling_rough}) is found \textit{if and only if} we have both $Y_2 \gg v^2$ and perturbativity. That is, assuming
eq.~(\ref{eq:decoupling_rough}) or assuming
$Y_2 \gg v^2$ and $Z_i/(4\pi) \lesssim \mathcal{O}(1)$ are just two equivalent ways to describe the
  same physical scenario corresponding to the decoupling limit. Eqs.~(\ref{eq:decoupling_def}) are yet another equivalent way, which specifies how the parameters of eq.~(\ref{eq:indep-real}) behave in that physical scenario.

This implies that the power-counting of the SMEFT and the HEFT matchings are going to be equivalent. It is true that, as shall be seen in detail, the SMEFT performs the expansion before SSB and the HEFT after it --- such that the former uses the Lagrangian parameter $Y_2$ as an expansion parameter, whereas the latter uses physical masses $m_{H}, m_{A}, m_{H^{+}}$. Yet, since those physical masses can be made large if and only if $Y_2 \gg v^2$ and $Z_i/(4\pi) \lesssim \mathcal{O}(1)$, the two expansions are the same, in the sense that they follow the same power-counting. Given the set of independent parameters of eq.~(\ref{eq:indep-real}), that power-counting is organized by powers of $\xi$, as defined in eq.~(\ref{eq:scaling}).

This leads to another aspect, related to the role of $c_{\beta-\alpha}$. The special scaling of $c_{\beta-\alpha}$ in eq.~(\ref{eq:scaling_of_camb}), as well as the subsequent $\xi$ power-counting introduced in eq.~(\ref{eq:scaling}), both follow from the choice of $c_{\beta-\alpha}$ as an independent parameter. If, instead of $c_{\beta-\alpha}$, one of the $Z_i$ were chosen as independent  --- say, $Z_6$ --- one would simply need to require $Z_6$ to obey $Z_6/(4\pi) \lesssim \mathcal{O}(1)$, in which case the expansion would simply be in inverse powers of $\Lambda^2$.%
\fn{Just as $c_{\beta-\alpha} \sim \mathcal{O} (v^2/\Lambda^2)$, the scaling $Z_6/(4\pi) \lesssim \mathcal{O}(1)$ would ensure not only perturbativity, but also that $m_h$ would be fixed. This last aspect can be seen by considering eq.~(\ref{eq:2HDM-masses-sq-var-a}), which is equivalent to eq.~(\ref{eq:2HDM-masses-sq-a}), but with $c_{\beta-\alpha}$ replaced by $Z_6$. It is clear that, as long as perturbativity is ensured (all $Z_i$ obeying $Z_i/(4\pi) \lesssim \mathcal{O}(1)$), the scenario of very large $Y_2$ will imply the cancellation of $Y_2$ in the expression~(\ref{eq:2HDM-masses-sq-var-a}).}
The two scenarios --- the one in which $c_{\beta-\alpha}$ is independent, and the one in which $Z_6$ is independent --- are perfectly equivalent. Note also that, if $Z_6$ were independent, we would find $\cba = Z_6 \, v^2/Y_2 + \mO(v^4/Y_2^2)\sim\mO(\xi^1)$, so that the scaling of the $\cba$ mixing in eq.~(\ref{eq:scaling}) would show up in a natural way.

Also relevant is an aspect concerning the mass states. As suggested above, the extreme case of the decoupling limit --- namely, $Y_2 \to \infty$ taken in a way consistent with perturbativity --- implies $c_{\beta-\alpha} \to 0$. This, in turn, implies $h_1^{\mathrm{H}} \to h$ and $h_2^{\mathrm{H}} \to -H$ 
by eq.~(\ref{eq:diagonalization}). It follows that $h_1^{\mathrm{H}}$ and $h_2^{\mathrm{H}}$ effectively correspond to mass states in the extreme decoupling. In the case in which $Y_2$ is very large but finite, there will be differences between $h_2^{\mathrm{H}}$ and the mass state $-H$ which are proportional to $\cba \sim \mathcal{O} (v^2/\Lambda^2)$.

Finally, we have been discussing how eq.~(\ref{eq:decoupling_rough}) can be obtained without spoiling perturbativity. We should keep in mind, however,  that the latter (perturbativity) is not restricted to that equation. Put another way, there are issues concerning perturbativity which are independent of the limit of heavy scalar masses. A simple example is provided by $\beta$; even though this parameter is independent of eq.~(\ref{eq:decoupling_rough}), its values can be such that  perturbativity is violated.%
\fn{For example, via the interactions of between $h$ and fermions (cf. e.g. ref.~\cite{Ferreira:2014naa}), or via $Z_2$ and $Z_7$ (cf. eqs.~(\ref{eq:theZs-real})). Finally, in some of the four types of 2HDM, $\beta$ can also cause a delayed decoupling~\cite{Haber:2000kq,Gunion:2002zf,Ferreira:2014naa}.}
Note that this feature is already present in the full 2HDM, so that it is not specific to an EFT expansion. This also means that the expansion of eq.~(\ref{eq:scaling}) does not ensure that perturbativity will be obeyed order by order in $\xi$; it only ensures that eq.~(\ref{eq:decoupling_rough}) can be obtained without violating perturbativity.

%% file: smeft.tex
\label{smeft}

As referred  to in the Introduction, the starting point of the SMEFT is the SM before SSB (to which higher-dimensional operators are added). 
Therefore, the SMEFT matching to the 2HDM must be done in such a way that the integration out of the heavy degrees of freedom of the 2HDM happens before SSB. Yet, here we are faced with a problem: not all the mass states of the 2HDM are defined before SSB. In fact, as seen above, the states $h_1^{\mathrm{H}}$ and $h_2^{\mathrm{H}}$ mix after SSB, and their mass matrix is diagonalized to yield the mass states $h$ and $H$. In that case, how can the heavy state $H$ be integrated out before SSB, if it is not even defined by then?

The answer has to do with the decoupling limit. We saw above that, in the extreme decoupling limit ($Y_2\rightarrow \infty$), $h_2^{\mathrm{H}}$ becomes a mass state. In that case, the doublet $H_2$ of the Higgs basis can be integrated out before SSB: on the one hand, the fact that $H_2$ is a doublet of ${\rm SU_L(2)}$ means that the states contained in it can be integrated out as a whole (without violating the symmetries of the theory before SSB). On the other hand, by eq.~(\ref{eq:Higgs_basis_param}), all its states become very heavy in that extreme decoupling scenario. 

$H_2$ is then integrated out at tree-level. This means a) assuming $H_2$ can be expressed as an expansion in inverse powers of $Y_2$, b) deriving a truncated solution for $H_2$ using equations of motion (EoM) and c) plugging that solution back in the original Lagrangian. The resulting Lagrangian will thus be itself an expansion in inverse powers of $Y_2$.%
\fn{Recall that we assumed the extreme decoupling scenario. Relaxing this assumption (i.e. taking $Y_2$ to be not so large) corresponds to considering higher powers in $1/Y_2$.}
This parameter is identified with the squared UV scale $\Lambda^2$, and the resulting EFT can be written in the format of the SMEFT. This exercise has been performed up to $\mathcal{O}(1/\Lambda^4)$ in ref.~\cite{Dawson:2022cmu}; in terms of the operators of dimension-6 and dimension-8 of the bases of refs.~\cite{Grzadkowski:2010es} and~\cite{Murphy:2020rsh}, respectively, the result reads:
\be
\label{eq:SMEFT-Lag}
\mathcal{L}_{\mathrm{SMEFT}} = \mathcal{L}_{\mathrm{SM}}
+
\dfrac{C_{\mathcal{H}}}{\Lambda^2} (\mathcal{H}^{\dagger} \mathcal{H})^3
+
\dfrac{C_{\mathcal{H}^8}}{\Lambda^4} (\mathcal{H}^{\dagger} \mathcal{H})^4
+
\dfrac{C_{\mathcal{H}^6}^{(1)}}{\Lambda^4} (\mathcal{H}^{\dagger} \mathcal{H})^2 \left(D_{\mu} \mathcal{H}\right)^{\dagger} \left(D^{\mu} \mathcal{H}\right) + \, ... \, + \mathcal{O}(1/\Lambda^6),
\ee
where $\mathcal{L}_{\mathrm{SM}}$ is the SM Lagrangian, $\mathcal{H}$ is the  Higgs doublet of the SMEFT expansion and the ellipses represent terms with fermions.%
\fn{Again, fermions are not relevant for our purposes. $\mathcal{H}$ is related to $H_1$ by a normalization factor; cf. ref.~\cite{Dawson:2022cmu} for details.}
The expressions for the WCs read~\cite{Dawson:2022cmu}:
\bs
\label{eq:matching_original}
\bea
\dfrac{C_{\mathcal{H}}}{\Lambda^2} &=& \dfrac{Z_{6}^2}{\Lambda^2} + \dfrac{2}{\Lambda^4} \Big( Y_3 Z_{1} Z_{6} - Y_3 Z_{345} Z_{6}
+ Y_1 Z_{6}^2 \Big), \\
\dfrac{C_{\mathcal{H}^8}}{\Lambda^4} &=& \dfrac{1}{\Lambda^4} \Big(2 Z_1 Z_{6}^2 - Z_{345} {Z_{6}}^2\Big), \\
\dfrac{C_{\mathcal{H}^6}^{(1)}}{\Lambda^4} &=& - \dfrac{{Z_{6}}^2}{\Lambda^4}.
\eea
\es
We can rewrite these matching relations in terms of the parameters of eq.~(\ref{eq:indep-real}). To that end, we use eqs.~(\ref{eq:theZs-real}) and, after assuming eq.~(\ref{eq:decoupling_def_a}), we consider the scaling of eq.~(\ref{eq:scaling}) and expand up to $\mathcal{O}(\xi^2)$. The result is:%
\bs
\label{eq:WCs-after-decoupling}
\bea
\label{eq:cH}
\dfrac{C_{\mathcal{H}}}{\Lambda^2}
&=&
c_{\beta-\alpha}^2 \,  (\sqrt{2} G_F)^2  \Big[{\Lambda}^2 - 4 \, (m_{h}^2 - \Delta m_H^2) \Big], \\
\dfrac{C_{\mathcal{H}^8}}{\Lambda^4} &=& 2 \, c_{\beta-\alpha}^2 \, (\sqrt{2} G_F)^3 \, \left(m_h^2 - \Delta m_{H}^2\right), \\
\dfrac{C_{\mathcal{H}^6}^{(1)}}{\Lambda^4} &=& - c_{\beta-\alpha}^2 \, (\sqrt{2} G_F)^2 ,
\eea
\es
where $G_F$ is the Fermi constant.%
\fn{Ref.~\cite{Dawson:2022cmu} considered the scenario in which $\Delta m_{H} = \Delta m_{A} = \Delta m_{H^+} = 0$, which is stronger than what is required by the decoupling limit of eqs.~(\ref{eq:decoupling_def}). We write the expressions in terms of $G_F$ instead of the vev, since the relation between the two gets corrections in the SMEFT; see ref.~\cite{Dawson:2022cmu} for details.}
The appearance of $\Lambda$ in the numerator of the right-hand side of eq.~(\ref{eq:cH}) is a consequence of our choice of $\cba$ as independent parameter.
Note also that there is no information about $\beta$ or asymmetry in $c_{\beta-\alpha}$ (i.e. odd powers of $c_{\beta-\alpha}$).
Finally, among the $\Delta m^2$ parameters introduced in eq.~(\ref{eq:decoupling_def_a}), only $\Delta m_H^2$ shows up, and always in the form $m_h^2 - \Delta m_H^2$.

%% file: heft_new.tex
\label{heft}

We saw in the Introduction that the HEFT considers the SM Higgs field $h$ to be a gauge singlet. This means that the HEFT matching to the 2HDM can only be accomplished if the heavy degrees of freedom of the 2HDM are integrated out after SSB. Contrary to the SMEFT approach, then, the HEFT  matching to the 2HDM starts with physical states, i.e. states with well defined masses, without mixing terms in the propagator. 
As a result, one can directly 
integrate out  the heavy mass states $H$, $A$ and $H^\pm$.   
Consistent with the discussion of section~\ref{perturbativity}, however, we will show that such an operation cannot be done by considering an expansion simply in inverse powers of $m_{H}$, $m_A$ and $m_{H^+}$. More than that, the scaling of $c_{\beta-\alpha}$ must be taken into account, or else there will be no consistent 
decoupling, since perturbativity is lost.

This danger can be illustrated by considering the cubic self-interaction of $h$. As with any three-point function, this interaction is not affected by the integration out of the heavy states at tree-level.%
\footnote{
The reason is that the solution of the EoM for a given heavy particle will always depend on terms which contain at least two light fields (since there are no bilinear terms in the Lagrangian which depend on two different fields, by definition of mass eigenstates). This means that, when replacing this heavy-particle EoM solution in the original UV Lagrangian, 
the two-point functions containing the heavy field will yield effective operators with four or more light fields. UV interaction terms with only one heavy field must also contain at least another two light fields and, hence, the corresponding effective operator has at least four light particles when the heavy scalar EoM solution is substituted. The same thing happens for UV interaction terms with two or more heavy fields, which give place to low-energy operators with four or more light fields for identical reasons~\cite{Henning:2014wua}. 
}
Thus, the cubic self-interaction of $h$ of the HEFT Lagrangian is obtained by considering the same interaction in the 2HDM Lagrangian and simply applying the EFT expansion. The 
Feynman rule for the cubic self-interaction of $h$ in the 2HDM reads:
\bea
\label{eq:triple-Higgs}
&& 
\frac{3 i
\csc^2(2 \beta)}{2 v} \Bigg\{ 
s_{\beta-\alpha} 
\cos (4
\beta ) \Big[-3 c_{\beta-\alpha}^4 m_{H}^2-2 c_{\beta-\alpha}^2
Y_2 +\left(3 c_{\beta-\alpha}^4+c_{\beta-\alpha}^2+1\right)
m_{h}^2\Big] \nonumber\\ 
&& \hspace{25mm} + c_{\beta-\alpha}^3 \sin (4 \beta ) \Big[\left(1-3
c_{\beta-\alpha}^2\right) m_{h}^2 + \left(3 c_{\beta-\alpha}^2-2\right)
m_{H}^2+2 Y_2\Big]
\nonumber \\
&& \hspace{25mm} + s_{\beta-\alpha} 
\Big[2 c_{\beta-\alpha}^2
Y_2 - c_{\beta-\alpha}^4 m_{H}^2 + \left(c_{\beta-\alpha}^4-c_{\beta-\alpha}^2-1\right)
m_{h}^2\Big]\Bigg\}  \, .
\eea
with $s_{\beta-\alpha}=\sqrt{1-c_{\beta-\alpha}^2}$. 
From this expression, we realize that an EFT expansion that considers simply inverse powers of $m_{H}$, $m_A$ and $m_{H^+}$ is doomed to inconsistency. This is because eq.~(\ref{eq:triple-Higgs}) contains 
positive powers of those heavy masses, so that the final HEFT Lagrangian can never be simply an expansion in inverse powers of those masses.%
\fn{This is also true for the quartic self-interaction of $h$. As discussed in  Appendix~\ref{singlet-appendix}, the singlet model of ref.~\cite{Buchalla:2016bse} is very special, since the cubic light-Higgs interactions does not scale with the heavy mass.}
Note that this has physical consequences, since observables like $WW \to hh$ would suffer the same inconsistency. 
On the other hand, 
an expansion according to the 
$\xi$-scaling in  
eqs.~(\ref{eq:decoupling_def}) and~(\ref{eq:scaling}) leads to a well-behaved cubic self-interaction of $h$.

The conclusion is then clear: the HEFT Lagrangian cannot be obtained from 
the 2HDM simply by performing an expansion in inverse powers of the heavy masses. 
 Decoupling and perturbativity in the UV theory require the consistent scaling in eqs.~(\ref{eq:decoupling_def}) and~(\ref{eq:scaling}),  
which leads to a well defined expansion.
The heavy states $H$, $A$ and $H^+$ can then be integrated out.
As mentioned above, this operation cannot affect three-point functions, which are thus trivially derived from the equivalent function in the UV model simply by applying eq.~(\ref{eq:decoupling_def_a}) and expanding according to eq.~(\ref{eq:scaling}). By contrast, 
vertices with four particles or more  receive contributions from  integrating out the heavy states. 

 It is to this procedure that we now turn.
To that end, and as described in the Introduction, we treat $h$ and the $\omega^a$ separately, such that the latter are embedded into a unitary matrix $U$.
The scalar doublets of the Higgs basis then take the form:%
\footnote{The inclusion of the $U$ matrix in the second doublet $H_2$ removes the Goldstone bosons from the potential in eq.~(\ref{eq:potential}). This was also noted in Ref.~\cite{Ciafaloni:1996ur}, but a different parametrization was used to eliminate the problem.}
\begin{align}
\mathcal{H}_1 = \frac{v+h_1^H}{\sqrt{2}} \ U(\omega)\, 
\begin{pmatrix}
0 \\
1
\end{pmatrix},
\hspace{3mm}
\mathcal{H}_2 = U(\omega)\,   
\begin{pmatrix}
H^+ \\
\frac{1}{\sqrt{2}}(h_2^{\mathrm{H}} + i A)
\end{pmatrix}\,.
\label{eq:sph-param}
\end{align}
We choose the unitary gauge, where the Goldstone bosons are eliminated from the theory, i.e. $U=1$ (our results were checked in an arbitrary $R_\xi$ gauge).
This has the advantage that there are no interactions with more than four fields.%
\fn{Alternative parameterizations of $U$ are common, such as the
spherical one ( 
$U= \sqrt{1- \omega^a \omega^a/v^2} + i \omega^a \sigma^a/v\, $) or the exponential one ($U=\exp\{i \omega^a \sigma^a /v\}$). 
 In general, 
one would need to expand $U$ up to the desired order.}   
Following ref.~\cite{Buchalla:2016bse}, we write the terms of $\mathcal{L}_{\rm 2HDM}$ involving scalars in such a way that we isolate the heavy scalars:
\begin{eqnarray}
\mL_{\rm 2HDM} &\supset& 
\frac{1}{2}(\partial_\mu H^a)^2  -\frac{1}{2} (M^2)^{ab} H^a H^b   
 + J_0 + J_1^a H^a
\nonumber\\
&& +J_2^{ab} H^a H^b +J_3^{abc} H^a H^b H^c + J_4^{abcd} H^aH^b H^c H^d\, ,
\label{eq:L2hdmjs}
\end{eqnarray}
where $(M^2)^{ab}$ is a diagonal matrix, $H^a=(H,\, A, \, H_3,\, H_4)$, with $H^\pm\equiv(H_3\mp i H_4)/\sqrt{2}$, and the $J_k$ contain only light fields.%
\fn{In particular, the part of the Lagrangian without heavy fields is encoded in $J_0$. Note that the derivation could also be done for $H^+$ and $H^-$ (instead of $H_3$ and $H_4$), but the expressions would not be as symmetric and simple as those presented here. 
The two bases are related by $H_3 J^{H_3} + H_4 J^{H_4}=H^+ J^{H^-} + H^- J^{H^+}$, with $J^{H^{\pm}}= (J^{H_3} \mp i J^{H_4})/\sqrt{2}$.
Also note that the Lagrangian terms quadratic in $H^a$ have been split in the form: terms without light fields are provided by $\frac{1}{2}(\partial_\mu H^a)^2-\frac{1}{2}(M^2)^{ab}H^aH^b$; terms with light fields have been placed in $J_2^{ab} H^a H^b$.}
The expressions for the $J_k$ are given in Appendix~\ref{2hdm-appendix}.

Each heavy scalar $H^a$ is integrated out at tree-level by solving its EoM:
\begin{eqnarray}
\label{eq:eom2hdm}
J_1^a\, +\, (-\partial^2-M^2+2 J_2)^{ab} H^b + 3 J_3^{abc} H^b H^c +4 J_4^{abcd} H^b H^c H^d \,=\, 0\, .
\end{eqnarray}
As mentioned before, the auxiliary parameter $\xi$ will act as the \textit{de facto} parameter of the expansion, as in the SMEFT case.  This means that eq.~(\ref{eq:eom2hdm}) will be solved iteratively in powers of $\xi$, so that the solution for the heavy fields will itself be given as a series in $\xi$.
As can be anticipated, 
even the lowest orders contain a long a list of
terms. We present only those which are relevant for the tree-level scattering processes we are interested in: $WW \rightarrow hh$ and $hh\rightarrow hh$.\fn{Since the process $ZZ \rightarrow hh$ would allow us to find the same matching as $WW \rightarrow hh$, and the comparison between the 2HDM and the EFT yield similar results for both processes, we have chosen $WW \rightarrow hh$ to assess the accuracy of the EFT fit.}  Other processes such as $WW \rightarrow WW$ depend solely on one EFT parameter, i.e the $a$ HEFT coupling that does not receive a modification from integrating out a heavy field at lowest order in HEFT and is the same as in the 2HDM. This is in contrast to $WW \rightarrow hh$ and $hh\rightarrow hh$ that involve corrections to $b$ and $d_4$, respectively. For this reason we will focus on these processes for our comparison.   
\fn{The general solution for $H$ and $H^+$ (containing all terms up to $\mathcal{O}(\xi^3)$, up to interactions with four particles) will be provided as supporting material with this manuscript.}
The heavy state $A$ does not play any role in these scatterings, so that it will be ignored in the following.
We then have:%
\be
\label{eq:heavysolutions}
H = \sum_{i=0}^{\infty} H_{(\xi^i)},
\qquad
H^+ = \sum_{i=0}^{\infty} H^+_{(\xi^i)}\, ,
\ee
where 
the lowest orders are,
\bs
\label{eq:solutionsheavyfieldsb}
\bea
\label{eq:solheavyxi0b}
H_{(\xi^0)} &= & H^+_{(\xi^0)} = 0 , \\
\label{eq:solheavyxi1b}
H_{(\xi^1)} & = & -\frac{3 c_{\beta-\alpha} h^2 }{2 v} ,\\
H^+_{(\xi^1)} & = & 0,\\
\label{eq:solheavyxi2b}
H_{(\xi^2)} & =  &\frac{2c_{\beta-\alpha}}{v \Lambda^2} \Delta m_{H}^2 h^2+\frac{
2 c_{\beta-\alpha}}{v  \Lambda^2} m_W^2  W_{\mu } W^{\dagger \mu }
 +
 \frac{3c_{\beta-\alpha}}{v \Lambda^2}  
 \left[(\partial^\mu h \partial_{\mu }  
 h) +h\left(\partial^2 \ h\right)\right]
 , \\
 H^+_{(\xi^2)} & =  & 
 -\frac{i c_{\alpha-\beta} M_W }{2 v \Lambda^2}  \left[h \left(\partial _{\mu }W^{\mu }\right)+2 W_{\mu } \left(\partial^{\mu }h\right) \right], \\
H_{(\xi^3)} &= & \frac{c_{\beta-\alpha} h^2 }{4 t_{\beta}^2 v \Lambda^4}  \left[c_{\beta-\alpha}^2 \left(3 t_{\beta}^4-2 t_{\beta}^2+3\right) \Lambda^4
-3 c_{\beta-\alpha} \left(t_{\beta}^2-1\right) t_{\beta} \Lambda^2  2 (2
   \Delta m_{H}^2-m_{h}^2)-8 {\Delta m_{H}^2}^2 t_{\beta}^2\right]
\nonumber \\
& & -\frac{7 \Delta m_{H}^2 c_{\beta -\alpha }} {\Lambda ^4 v}  \left[\left(\partial^\mu h \right) \left( \partial_\mu h\right)+h \left(\partial^2 h\right)\right]
   -\frac{3 c_{\beta -\alpha }}{\Lambda ^4 v} \left[\left(\partial^2 h\right) \left( \partial^2 h\right)+h \left(\partial^2 \partial^2 h\right) \right]  \nonumber\\
   & & -\frac{6 c_{\beta -\alpha } }{\Lambda ^4 v} \left[(\partial^\mu h \partial^\nu h)(\partial_\mu h \partial_\nu h)+\right(\partial _{\mu }h \left) (\partial^\mu \partial^2 h)+\left(\partial _{\nu }h \right) (\partial^2 \partial^\nu h)\right] \nonumber
   \\
   &  & -\frac{2 c_{\beta-\alpha} m_W^2}{ v \Lambda^4}   \left[ 
 W^{\dagger \nu } \left(\Delta m_{H}^2 W_{\nu }+\partial^2 W_{\nu }\right)+2 \left(\partial^{\mu } W^{\nu} \right) \left( \partial_{\mu } W_{\nu }^{\dagger }\right)+  W_{\nu } \left(\partial^2 W^{\dagger \nu }\right)   \right]\, , 
    \\
\label{eq:solutionsheavyfieldsb-3+}
   H^+_{(\xi^3)} &= & -\frac{i m_W c_{\beta -\alpha }}{v \Lambda ^4}  \left[h (\partial ^2\partial ^{\nu }W_{\nu })+(\partial^2 h)
   (\partial^{\nu }W_{\nu })\right]
   \nonumber \\
   & & + \frac{i c_{\beta-\alpha} m_W}{v \Lambda ^4}  \left[h \left(\partial^2\partial ^{\nu }W_{\nu }\right)+2 W^{\nu } \left(\partial^2\partial _{\nu }h\right)\right]-\frac{4 i m_W c_{\beta -\alpha }}{v \Lambda^4} \left(\partial ^{\mu }\partial ^{\nu }h\right) \left(\partial _{\mu }W_{\nu }\right)
   \nonumber \\
   && - \frac{i c_{\beta-\alpha} \Delta m_{H_{+}}^2  m_W}{v \Lambda^4} \left[h \left(\partial ^{\nu }W_{\nu }\right)+2 W^{\nu } \left(\partial _{\nu }h\right)\right].    
\eea
\es

Note that the first non-vanishing solution starts at $\mO(\xi)$ for $H$, and at $\mO(\xi^2)$ for $H^+$. As a consequence, the integration out of $H$ and $H^+$ will contribute to $WW \rightarrow hh$ at order $\mO(\xi^2)$ and $\mO(\xi^3)$, respectively.  Finally, contrary to what was done in the SMEFT, we performed the expansion in the HEFT up to $\mathcal{O}(\xi^3)$. We justify this difference of truncations between the SMEFT and the HEFT matchings at the end of this section.

By substituting the solutions for the heavy fields of eq.~(\ref{eq:heavysolutions}) in $\mathcal{L}_{\rm 2HDM}$, we obtain the effective HEFT Lagrangian.%
\fn{This contains in general terms that cannot be written in the form of eqs.~(\ref{eq:heftdef}) and~(\ref{eq:flare}), since they would require terms in the HEFT Lagrangian 
with additional derivatives. In the expressions for the HEFT matching in this paper, we will not be presenting such terms.
\label{fn:aux}}
Note that, for the two-to-two tree-level scatterings discussed in this article ($WW\to hh$ and $hh\to hh$), only the first line in eq.~(\ref{eq:L2hdmjs}) is required --- at any order in $\xi$. Since the $H^a$ EoM solutions contain at least two light fields, the effective operators in the second line of eq.~(\ref{eq:L2hdmjs}) will contain  five or more light fields.
Comparing the effective HEFT Lagrangian with that of eqs. ~(\ref{eq:heftdef}) and~(\ref{eq:flare}) results in the following matching equations:
\bs
\begin{eqnarray} 
\Delta a^2 \equiv a^2 \, -\, 1 &=& - c_{\beta-\alpha}^2  \, ,
\label{eq:a2-HEFT}\\
\Delta b\equiv b \, -\, 1&=&  - \, 3 \cba^2  + 4 \cba^2 \frac{\Delta m_{H}^2}{\Lambda^2} 
\label{eq:b-HEFT} 
\, +\, \mO(\xi^4)\, ,
\\
\Delta d_3\equiv d_3 -1 &=& - 2 \cba^2 \frac{\Lambda^2}{m_h^2} \, +\, \frac{1}{2}\cba^2    
\label{eq:l3-HEFT} \\
&& + \cba^3 \bigg[ -
\cot(2\beta)   \left(1-\frac{2  \Delta m_{H}^2}{m_h^2}\right) +2 c_{\beta -\alpha } \cot^2(2\beta) \frac{  \Lambda^2
  }{m_h^2} \bigg]  
   +\mO(\xi^4)\, ,  \nonumber \\
\Delta d_4\equiv d_4 -1 &=& - 12 \cba^2 \frac{\Lambda^2}{m_h^2} 
\, +\, \cba^2  \left(\frac{16\Delta m_{H}^2}{m_h^2}-11\right)  
\label{eq:l4-HEFT} \\
&& + \cba^2 \bigg[ 
2  c_{\beta -\alpha }^2\frac{ \Lambda^2}{m_h^2} \left(22 \cot^2(2\beta)-17\right) 
- 22 c_{\beta -\alpha }  \cot(2\beta)
    \left(1- \frac{2 \Delta m_{H}^2}{m_h^2}\right) 
\nonumber\\&&\qquad\qquad 
    + 16\frac{ \Delta m_{H}^2}{\Lambda^2}
   \left(\frac{2- \Delta m_{H}^2}{m_h^2}   \right)
\bigg]  
   +\mO(\xi^4)\, .  \nonumber
\\ \nonumber
\end{eqnarray}
\es
In order to more easily compare  with the SM, we introduced the quantities with $\Delta$;
from eqs.~(\ref{eq:decoupling_def}) and~(\ref{eq:scaling}), it is easy to see that the SM limit ($\Delta a^2=\Delta b=\Delta d_3=\Delta d_4=0$) is recovered
at $\mathcal{O}(\xi^0)$.
For both $a^2$ and $b$,
the first deviation from the SM occurs at $\mO(\xi^2)$; this happens in such a way that $a^2$ has no additional contributions.
Note also that the $\beta$ dependence appears for the first time at 
$\mO(\xi^3)$ for $\Delta d_3$ and $\Delta d_4$ (second lines of eqs.~(\ref{eq:l3-HEFT}) and~(\ref{eq:l4-HEFT}), respectively).
Finally, the factors $\cot(2\beta)=(1-\tan^2\beta)/(2\tan{\beta})$ become large for $\beta \sim 0$ or $\beta\sim \pi/2$ (i.e., when $\tan{\beta}\to 0$ or $\infty$, respectively), and vanish for $\theta=\pi/4$ (i.e., when $\tan{\beta}=1$). 

We can compare the analytical results obtained in this section with the ones from SMEFT.
We start by realizing that, up to $\mathcal{O}(\xi^2)$, and just as in the SMEFT matching, there is no information about $\beta$ or 
odd powers
in $c_{\beta-\alpha}$, and the only $\Delta m^2$ parameter present is $\Delta m_H^2$.
We also observe that the relation found between eqs.~(\ref{eq:a2-HEFT}) and~(\ref{eq:b-HEFT}), $\Delta b = 3 \Delta a^2 + \mathcal{O}(\xi^3)$, is not compatible with the usual dimension-6 SMEFT constraint $
\Delta b \,=\, 2 \Delta a^2$~\cite{Gomez-Ambrosio:2022qsi}.
That is,
the HEFT matching to the 2HDM cannot be described by means of  a SMEFT Lagrangian that starts with dimension-6 operators. On the other hand,
if one assumes that the contributions from the dimension-6 SMEFT operators to $a$ and $b$ vanish (as is indeed the case in the SMEFT matching to the 2HDM, see ref.~\cite{Dawson:2022cmu}),
one obtains a dimension-8 constraint, which is precisely $
\Delta b \,=\, 3 \Delta a^2$~\cite{Gomez-Ambrosio:2022qsi,Gomez-Ambrosio-in-prep}. 
Regarding the Higgs potential term, a SMEFT Lagrangian starting at dimension--6 also requires the relation 
$\Delta d_4 
= 3 \Delta d_3^2 - 2 \Delta a^2/3  $ 
between the HEFT couplings~\cite{Gomez-Ambrosio:2022qsi,Gomez-Ambrosio:2022why}, where the latter $\Delta a^2$ comes from a finite Higgs field redefinition.
It is easy to observe that the values of $\Delta d_3$ and $\Delta d_4$ in the 2HDM obey this relation at $\mO(\xi)$, as $\Delta a=0$ at that order.

We end this section with a remark about the difficulty of implementation of the two EFT approaches to the 2HDM. 
The HEFT approach is considerably simpler to implement than the SMEFT one for the processes considered here.
First of all, the higher orders terms in SMEFT in general contain the SM Higgs doublet, which contains the SM vev. This means that two-point functions are in general affected; in particular, kinetic terms and the relations between masses and Lagrangian parameters 
need to be redefined. In the HEFT approach, this never happens, since the integration out of the heavy states only affects four-point functions or higher, as discussed above.
This is related to a second advantage, which
is that
the three-point functions in the HEFT approach at tree-level are trivially obtained from the corresponding functions in the 2HDM, which is not the case in the SMEFT approach.
Finally, for the processes considered here, the HEFT approach at tree-level does not require the formal procedure of integrating out heavy states.
The same results can be obtained simply by considering the amplitudes of the 2HDM contributing to the process at stake, and applying the expansion of eqs.~(\ref{eq:decoupling_def}) and~(\ref{eq:scaling}) directly to them.
All of this allows us to easily derive the $\mathcal{O}(\xi^3)$ results in the HEFT expansion (Appendix~\ref{2hdm-appendix}). The derivation of the same order results in the SMEFT (which involve dimension-10 operators) is beyond the scope of this work.

%% file: results.tex
\label{results}

We now turn to our numerical results. We assume 
that $H$, $A$ and $H^+$ are all degenerate and  we define the quantity $\Delta \Lambda$, such that:%
\footnote{For the processes considered here, $A$ does not play any role, so that the results are independent of $m_A$.}
\be
\label{eq:def-dl}
m_H = m_A = m_{H^+} = \Lambda + \Delta \Lambda.
\ee
Comparing with eqs.~(\ref{eq:2HDM-masses-sq-c}) and~(\ref{eq:2HDM-masses-sq-d}), and recalling that $Y_2 = \Lambda^2$ (eq.~(\ref{eq:decoupling_def_a})), we realize that $\Delta \Lambda$ measures the amount of mass in $m_A$ and $m_{H^+}$ which is not generated by the Lagrangian parameter $Y_2$. In other words, $\Delta \Lambda = 0$ implies that $m_A$ and $m_{H^+}$ are entirely generated by $Y_2$, whereas larger and larger values of $\Delta \Lambda$ imply larger and larger contributions from the vev.%
\fn{Negative values of $\Delta \Lambda$ are in principle also possible, but they are generally ruled out by theoretical constraints.
}
Eq.~(\ref{eq:def-dl}) implies that the quantities defined in eq.~(\ref{eq:decoupling_def_a}) obey:
\be
\label{eq:dl-scaling}
\Delta m_H^2 = \Delta m_A^2 = \Delta m_{H^+}^2 = 2 \, \Lambda \, \Delta \Lambda + (\Delta \Lambda)^2.
\ee
Accordingly, the new parameter scales as $\Delta \Lambda \sim \mathcal{O}(v^2/\Lambda) \sim \mO(\xi^{1/2})$.

Naively,  $\Delta \Lambda$ is expected to control the increase of accuracy of the HEFT matching over the SMEFT one. The reason is that the heavy mass parameter in the SMEFT matching is $Y_2$ (which is set equal to $\Lambda^2$), whereas in the HEFT the heavy mass parameters are the heavy masses (which are given by eq.~(\ref{eq:def-dl})). The HEFT thus contains information about $\Delta \Lambda$, so that, for large values of $\Delta \Lambda$, the agreement of the HEFT matching to the 2HDM is expected to be better than that of the SMEFT matching. A similar reasoning motivated the v-improved matching proposed in ref.~\cite{Brehmer:2015rna}.

However, two aspects should not be neglected. First, the numerators of the expressions of the SMEFT matching to the 2HDM in general depend on the masses. Therefore, they will in general depend on $\Delta \Lambda$ (and they indeed do: see eqs.~(\ref{eq:WCs-after-decoupling})).
Second, even if the HEFT heavy mass parameters are the heavy masses of the 2HDM, these are constrained to follow eqs.~(\ref{eq:decoupling_def}). It follows that the scaling $\Delta \Lambda \sim \mathcal{O}(v^2/\Lambda)$ implies a suppressed dependence of the HEFT matching on $\Delta \Lambda$. All in all, then, it is to be seen if a correlation exists between $\Delta \Lambda$ and an increase in accuracy of the HEFT matching over the SMEFT one.

\begin{figure}[htb!]
\centering
\includegraphics[width=0.7\textwidth]{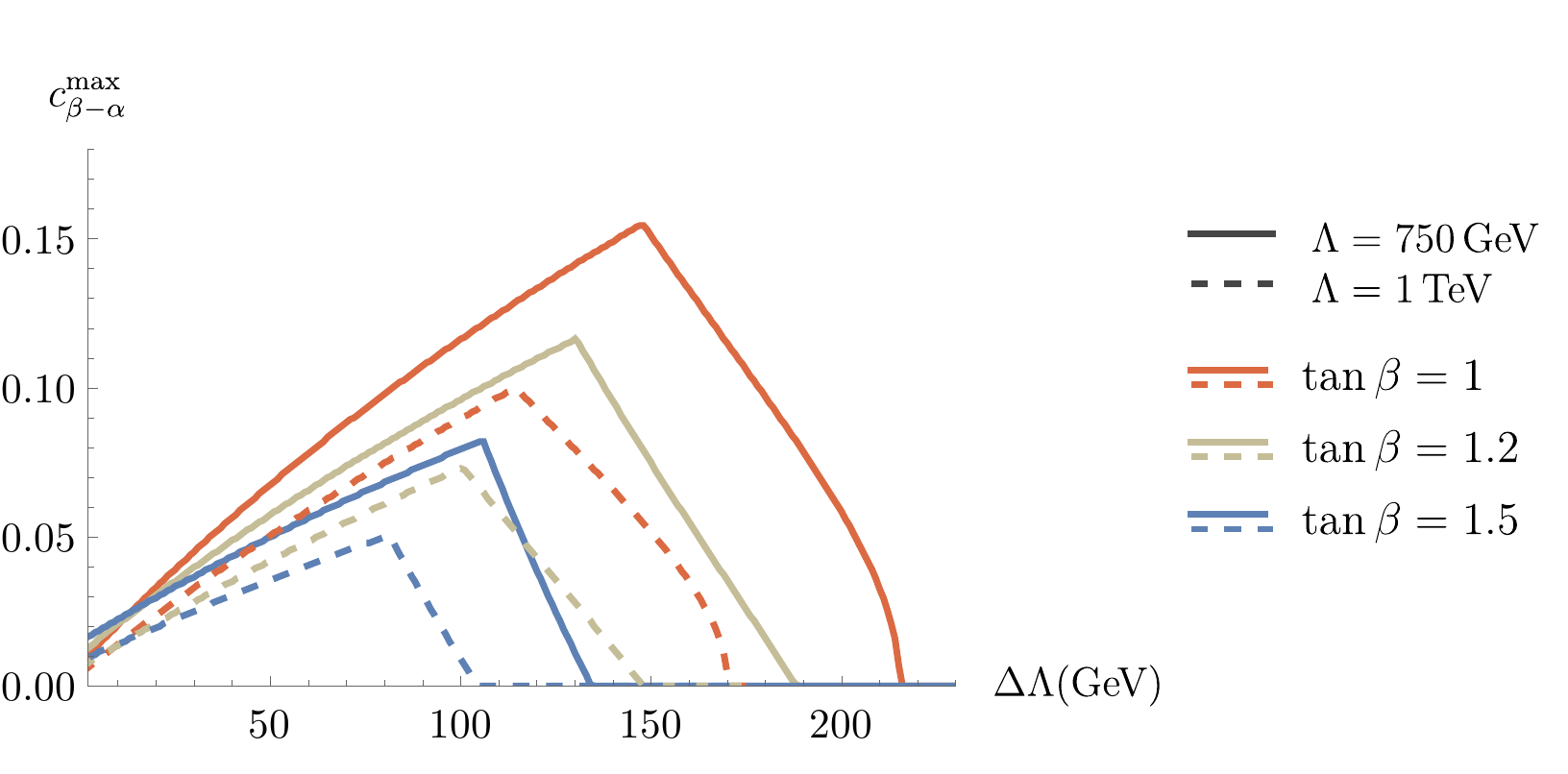}
\vspace{-3mm}
\caption{{\small Maximum value of $c_{\beta-\alpha}$ allowed by the theoretical constraints of the 2HDM, as a function of $\Delta \Lambda$. For each curve, the maximum value of $c_{\beta-\alpha}$ is determined by boundedness from below in the region where the curve has positive slope, and by perturbative unitarity in the region where the curve has negative slope.}}
\label{fig:TheoCons}
\end{figure}

For the numerical results that follow, we require our 2HDM results to comply with the theoretical constraints of perturbative unitarity and boundedness from below~\cite{Deshpande:1977rw,Kanemura:1993hm,Akeroyd:2000wc,Ginzburg:2005dt}, as well as electroweak precision measurements via the oblique parameters S, T and U~\cite{Branco:2011iw}. We start by ascertaining the relevance of these contraints on the parameter space. This can be seen in fig.~\ref{fig:TheoCons}, 
where the maximum value of $c_{\beta-\alpha}$ allowed ($c_{\beta-\alpha}^{\rm max}$) is shown versus $\Delta \Lambda$, for different values of $\Lambda$ and $\tan \beta$.
Each curve  shows an abrupt inflexion point; in all cases, the values of $\Delta \Lambda$ below that point are such that $c_{\beta-\alpha}^{\rm max}$ is determined by boundedness from below, whereas those above it have $c_{\beta-\alpha}^{\rm max}$ determined by perturbative unitarity.%
\fn{Given the assumed degeneracy of $H$, $A$ and $H^+$, the oblique parameters play no relevant role in our analyses. 
Boundedness from below requires that none of the elements of specific combinations of quartic parameters of the potential (usually in the original basis of the doublets $\Phi_1, \Phi_2$) take negative values (see e.g. ref. \cite{Deshpande:1977rw}). For the values of $\tan \beta$ considered in the figure, the most important element is $\lambda_2$. When written in terms of the parameters involved in the plot, and when expanded to first order in $c_{\beta-\alpha}$, this quartic parameter is of the form $c_1 - c_2 \, c_{\beta-\alpha}$. Here, $c_1$ and $c_2$ are real numbers which depend on $\Delta \Lambda$ and which, for the values of $\Delta \Lambda$ involved, are both positive. The requirement that $c_1 - c_2 \, c_{\beta-\alpha}$ is non-negative thus imposes an upper limit on the value of $c_{\beta-\alpha}$. Moreover, $c_1$ turns out to grow with $\Delta \Lambda$ twice as quickly as $c_2$, which explains the linear character of the positive-slope branch of the curves.
As for the negative-slope branch, it is determined by perturbative unitarity, which requires all the elements of another combination of quartic parameters of the potential to be smaller than a certain limit. For the values at stake here, the decisive element is $|\frac{3 \,  \left( {\lambda_1} + {\lambda_2} \right)  + \sqrt{9 \,  \left( {\lambda_1} - {\lambda_2} \right) ^2 + 4 \,  \left( 2 \, {\lambda_3} + {\lambda_4} \right) ^2}}{2}|\le 8 \pi $. Just as before, we can write it in terms of the parameters involved in the plot and expand it to second order in $c_{\beta-\alpha}$, in which case it acquires the form $c_3 + c_4 \, c_{\beta-\alpha}^2\le 8 \pi$. Just as $c_1$ and $c_2$, also $c_3$ and $c_4$ are real positive numbers (in the range of values at stake), such that $c_4$ grows with $\Delta \Lambda$. This happens in such a way that, from a certain value of $\Delta \Lambda$, the maximum allowed value of $c_{\beta-\alpha}$ is no longer determined by boundedness from below, but rather from perturbative unitarity. The inflexion point in each curve (where the negative-slope and the positive-slope unite) is thus a non-trivial combination of these two theoretical constraints.}
The figure also shows that the window of allowed values of $c_{\beta-\alpha}$ becomes narrower with both increasing $\Lambda$ and increasing $\tan \beta$. We checked, in particular, that scenarios with $\tan \beta \sim 1$ and $\Lambda \gg 1$ TeV have an extremely narrow allowed window, as do also scenarios with $\Lambda \sim 1$ TeV and $\tan \beta \gg 1$.
Finally, for the (large) values of $\Lambda$ shown, the largest value of $c_{\beta-\alpha}^{\rm max}$ allowed is around 0.15. The result is that one is restricted to be very close to the exact alignment limit $c_{\beta-\alpha} = 0$. Still, interesting results can be found inside that narrow window.

The 2HDM is  limited by numerous experimental results, of which the most stringent are Higgs coupling measurements, $b$ meson decays and searches for heavy Higgs bosons.  These limits depend on the couplings of the fermions to the Higgs doublets, and we assume that the couplings respect a $\mathbb{Z}_2$ symmetry.  The limits from Higgs couplings typically require that $c_{\beta-\alpha}$ be close to the alignment limit, and all of the values considered  below are currently allowed~\cite{CMS:2022dwd,atlas:2022vkf}.
The charged Higgs boson  that is present in the 2HDM contributes to the decay $b\rightarrow s\gamma$ and  current experimental results require that $\tan\beta > 1.2$~\cite{Haller:2018nnx}.  Additionally,
ATLAS and CMS have searched for heavy neutral scalars with the couplings of the 2HDM and for $\tan\beta > 1.2$,
the limit is quite weak, $m_H> 400$~GeV~\cite{ATLAS:2017tlw,CMS:2019bnu}.
 In the following, we shall take $\tan \beta = 1.2$ since, from figure~\ref{fig:TheoCons}, this gives the
 largest theoretically allowed region that is consistent with experiment. The results that follow were obtained independently via \textsc{FeynMaster}~\cite{Fontes:2019wqh,Fontes:2021iue} (and its accompanying software~\cite{Christensen:2008py,Alloul:2013bka,Nogueira:1991ex,Mertig:1990an,Shtabovenko:2016sxi,Shtabovenko:2020gxv}) and \textsc{FeynArts}~\cite{Hahn:2000kx}.

Before considering our numerical analysis of the SMEFT and the HEFT matchings to the 2HDM, we highlight that both approaches end up using the same expansion (in powers of $\xi$, defined in eq.~(\ref{eq:scaling})), since the decoupling limit of eq.~(\ref{eq:decoupling_def}) needs to be obeyed by both in order to have a weakly interacting perturbative 2HDM. Hence, even if they are structurally different --- the SMEFT matching complying with the symmetries of the SM before SSB, the HEFT one with those after SSB --- some of their results are very similar.
For example, both the three-point tree-level interactions between $h$ and fermions and between $h$ and gauge bosons are exactly the same in the two effective Lagrangians at $\mathcal{O}(\xi^2)$. This implies, in particular, that the fits to global Higgs signal strengths performed in ref.~\cite{Dawson:2021xei} will be the same in the SMEFT and in the HEFT matchings at that order.%
\fn{We refer to the fits which do not include the effects of the Higgs trilinear coupling, fig.~6 of ref.~\cite{Dawson:2021xei}. Note that even one-loop processes such as $gg \to h$ or $h \to \gamma\gamma$ are the same in both EFT approaches (at $\mathcal{O}(\xi^2)$).}

It turns out that the tree-level scatterings $WW\rightarrow hh$ and $hh\rightarrow hh$ are also identical at $\mathcal{O}(\xi^2)$.
This result does not appear obvious to us, since the individual Feynman rules contributing to the processes are different at $\mathcal{O}(\xi^2)$. Specifically, the $h^3$ coupling --- which contributes to both $WW\rightarrow hh$ and $hh\rightarrow hh$ --- involves derivatives in the SMEFT matching (recall eq.~(\ref{eq:SMEFT-Lag})), whereas in the HEFT matching it does not (as can be seen by applying eqs.~(\ref{eq:decoupling_def}) and to eq.~(\ref{eq:scaling}) to eq.~(\ref{eq:SMEFT-Lag})). But the fact that the local 4-point interactions ($WWhh$ in $WW \to hh$, and $h^4$ in $hh \to hh$) are also different exactly compensates for the difference in $h^3$ to $\mathcal{O}(\xi^2)$.

Note that this conclusion holds even before the assumption of degenerate heavy masses, eq.~(\ref{eq:def-dl}). That it holds in the case of degenerate heavy masses implies that it holds for all $\Delta \Lambda$. In other words, the parameter $\Delta \Lambda$ is irrelevant to compare the SMEFT and the HEFT matchings in $WW\rightarrow hh$ and $hh\rightarrow hh$ at tree-level at $\mathcal{O}(\xi^2)$, since the two approaches are analytically identical. 
In the following, we refer to the two identical matchings at $\mathcal{O}(\xi^2)$ simply as the EFT matching, and we investigate how accurately it describes the 2HDM results.

\begin{figure}[htb!]
\centering
\includegraphics[width=0.8\textwidth]{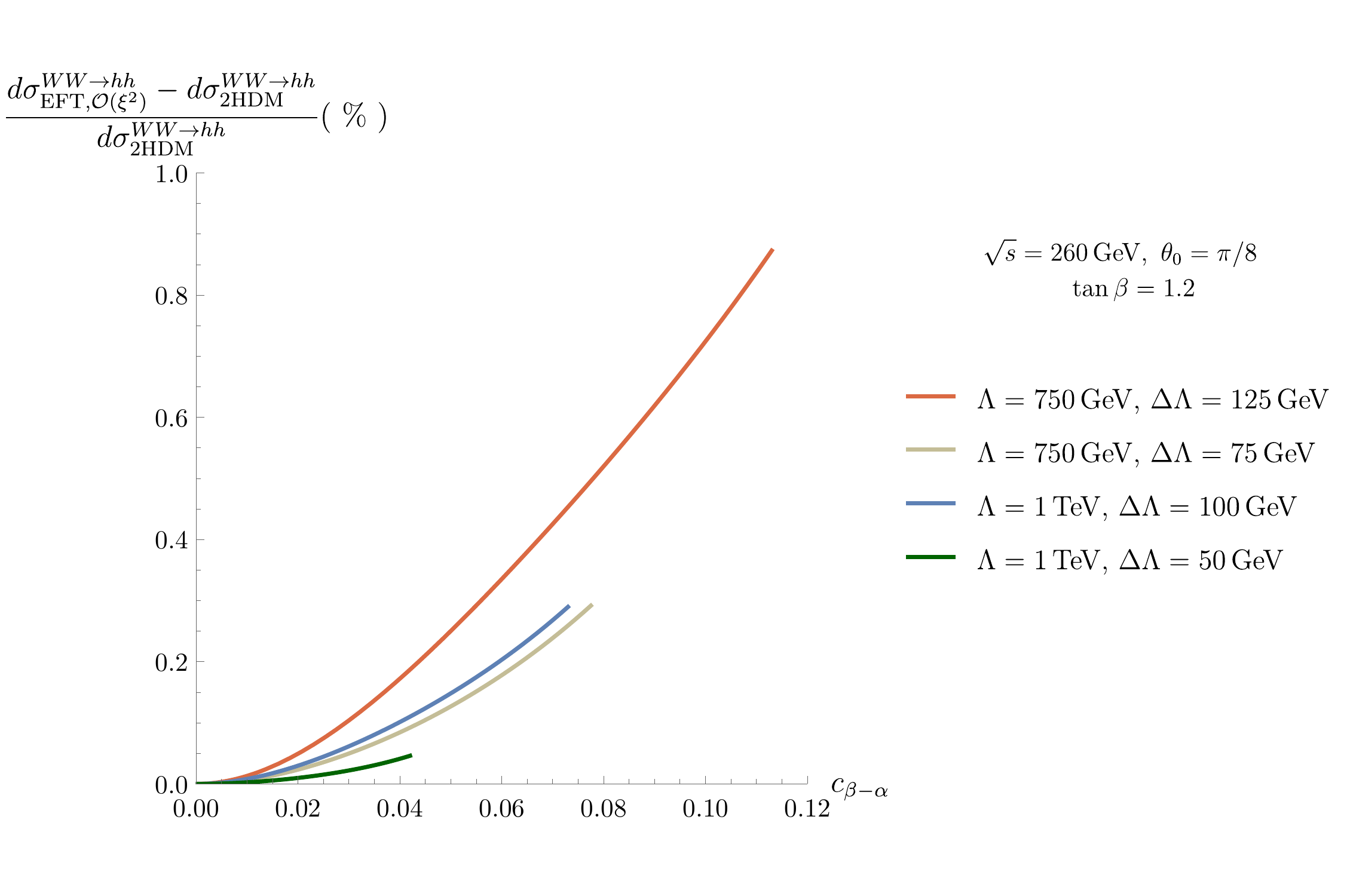}
\vspace{-10mm}
\caption{{\small Relative difference between the differential tree-level cross-sections for $WW \to hh$ in the 2HDM and in the EFT matching to the 2HDM at $\mathcal{O}(\xi^2)$ with $d\sigma\equiv {d\sigma\over d\theta}\mid_{\theta=\theta_0}$. 4 pairs of values of $\Lambda$ and $\Delta \Lambda$ are considered, according to the labels. For each curve, only the range of (positive values of) $c_{\beta-\alpha}$ allowed by the theoretical constraints is shown. All results assume a center-of-mass energy $\sqrt{s} = 260 \, {\rm GeV}$, a scattering angle $\theta_0=\pi/8$ and $\tan \beta = 1.2$.}}
\label{fig:WWtohh-SMEFT-vs-HEFT}
\end{figure}

We start by illustrating the case $WW \to hh$, depicted in figure~\ref{fig:WWtohh-SMEFT-vs-HEFT}. The plot shows the relative differential cross section 
between the 2HDM and the EFT matching at $\mathcal{O}(\xi^2)$, for different values of $\Lambda$ and $\Delta \Lambda$, and for a center-of-mass energy $\sqrt{s} = 260 \, {\rm GeV}$ and a scattering angle $\theta_0=\pi/8$.%
\fn{The general features of the plot are not sensitive to the specific values of $\sqrt{s}$ and $\theta_0$.
Moreover, the expressions for $d \sigma_{{\rm HEFT}, \mathcal{O}(\xi^2)}^{WW \to hh}$ and $d \sigma_{{\rm SMEFT}, {\mathcal{O}}(\xi^2)}^{WW \to hh}$ are consistently of $\mathcal{O}(\xi^2)$, in the sense that higher order effects resulting from squaring the amplitude were excluded.}
The plot only shows positive values of $c_{\beta-\alpha}$, and each curve is shown only up to the value of $c_{\beta-\alpha}$ where the theoretical constraints start being violated (cf. figure~\ref{fig:TheoCons}).
It is manifest that the EFT matching reproduces quite well the 2HDM, with relative differences smaller than $1\%$.

\begin{figure}[htb!]
\centering
\includegraphics[width=0.8\textwidth]{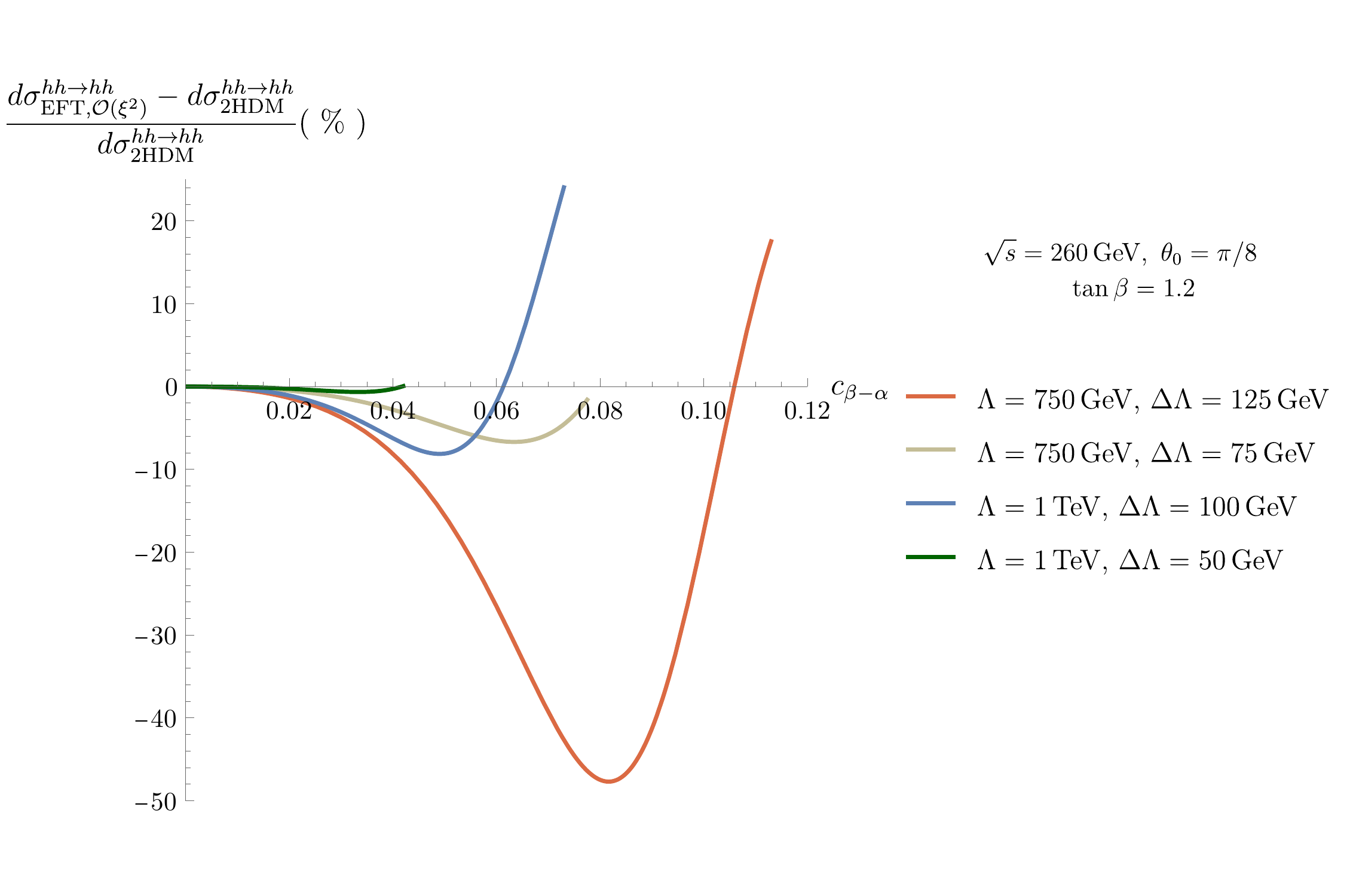}
\vspace{-10mm}
\caption{{\small The same as in figure~\ref{fig:WWtohh-SMEFT-vs-HEFT}, but for $hh \to hh$.}}
\label{fig:hhtohh-SMEFT-vs-HEFT}
\end{figure}

This is to be contrasted to what is shown in figure~\ref{fig:hhtohh-SMEFT-vs-HEFT}, which considers the same as in figure~\ref{fig:WWtohh-SMEFT-vs-HEFT}, but now for $hh \to hh$.
The EFT matching no longer faithfully reproduces the 2HDM result, allowing differences 
larger than $40\%$ for $\Lambda=750$ GeV, $\Delta \Lambda = 125$ GeV, $c_{\beta-\alpha} \sim 0.08$. These large values demonstrate that, in the region of parameter space considered,  $\mathcal{O}(\xi^2)$ is not enough in the EFT expansion. This means that, to accurately reproduce the 2HDM result, one would need a matching to dimension-10 operators in SMEFT, and to operators beyond the leading order in the derivative expansion in HEFT.

\begin{figure}[htb!]
\centering
\includegraphics[width=1\textwidth]{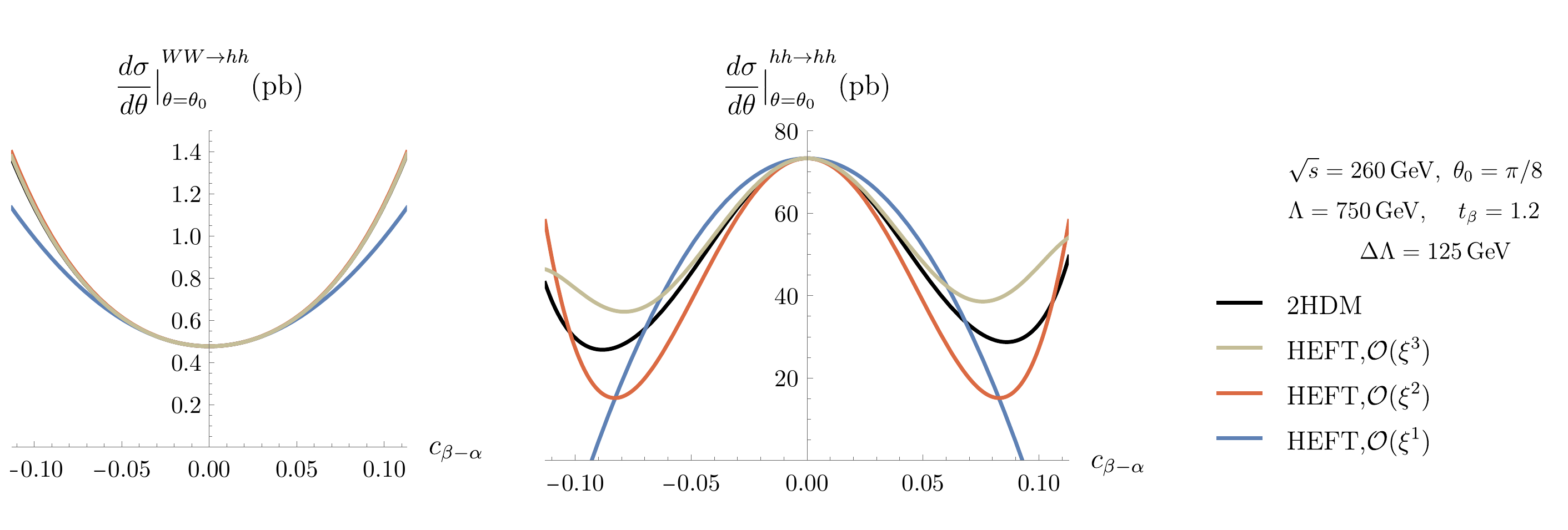}
\vspace{-10mm}
\caption{{\small Left: differential cross section for $WW \to hh$ at tree-level, both for the 2HDM (black), as well as for three different truncations of the HEFT matching (the black and the red are behind the beige).
Right: the same, but for $hh \to hh$. 
Both panels take $\sqrt{s}=260$ GeV, $\theta_0 = \pi/8, \Lambda = 750$ GeV, $\Delta \Lambda = 125$ GeV and $\tan \beta = 1.2$. The region of values of $c_{\beta-\alpha}$ shown is allowed by the theoretical constraints.}}
\label{fig:hhhh-HEFT-vs-2HDM}
\end{figure}

Figure~\ref{fig:hhhh-HEFT-vs-2HDM} displays again $WW \to hh$ and $hh \to hh$, but with three main differences: first, it shows the absolute values of the differential cross-sections; second, it includes negative values of $c_{\beta-\alpha}$; finally, it separately shows the different orders in the HEFT expansion, up to $\mathcal{O}(\xi^3)$.%
\fn{For the values of $t_{\beta}$, $\Lambda$ and $\Delta \Lambda$ considered, some values of $c_{\beta-\alpha}$ more negative than the ones shown in the plots are still allowed by theoretical constraints. Moreover, even if we are not showing all the terms $\mathcal{O}(\xi^3)$ in eqs.~(\ref{eq:solutionsheavyfieldsb}), we are including them in these plots. Finally, the $\mathcal{O}(\xi^1)$ curve yields negative values for $d \sigma^{hh \to hh}$ for $|c_{\beta-\alpha}| > 0.07$. These are unphysical (and thus not shown), and result from neglecting the higher order terms when taking the square of the amplitude.}
Several aspects are worth mentioning here. First, we stress that the plots show the HEFT matching, which we performed up to $\mathcal{O}(\xi^3)$, but which we are only assured of being identical to the SMEFT matching up to $\mathcal{O}(\xi^2)$.
Then, the right panel shows that the 2HDM result is slightly asymmetric in $c_{\beta-\alpha}$, even though the EFT matchings at $\mathcal{O}(\xi^2)$ do not contain this information, as discussed above.%
\fn{As we also noted, the EFT matching at $\mathcal{O}(\xi^2)$ does not have information about $\tan \beta$. This suggests that the two approaches will poorly reproduce the 2HDM whenever the latter shows a strong dependence on that parameter. On the other hand, and as discussed in the context of figure~\ref{fig:TheoCons}, a scenario with large $\Lambda$ and $\tan \beta$ significantly different from 1 will lead to the alignment limit $c_{\beta-\alpha}$, where the EFT matching coincides with the 2HDM.}

Concerning the different truncations, the right plot of figure~\ref{fig:hhhh-HEFT-vs-2HDM} illustrates that, while the lowest truncation in enough to reproduce the 2HDM for values of $c_{\beta-\alpha}$ very close to zero, the $\mathcal{O}(\xi^3)$ truncation is clearly the most appropriate one for the whole range of $c_{\beta-\alpha}$ shown. On the other hand, even that truncation is far from an exact reproduction of the 2HDM result, which indicates that the next order would be relevant. In other words, the convergence of the EFT expansion is quite slow for $hh \to hh$ for larger values of $c_{\beta-\alpha}$.
This is to be contrasted with the left panel, which shows the equivalent plot for $WW \to hh$. There, a faithful reproduction of the 2HDM results is obtained immediately at $\mathcal{O}(\xi^2)$, in which case higher orders are not needed. Nevertheless, both panels also show that, again for larger values of $c_{\beta-\alpha}$, the $\mathcal{O}(\xi^1)$ truncation clearly fails to reproduce the UV model.

\begin{figure}[htb!]
\centering
\includegraphics[width=1\textwidth]{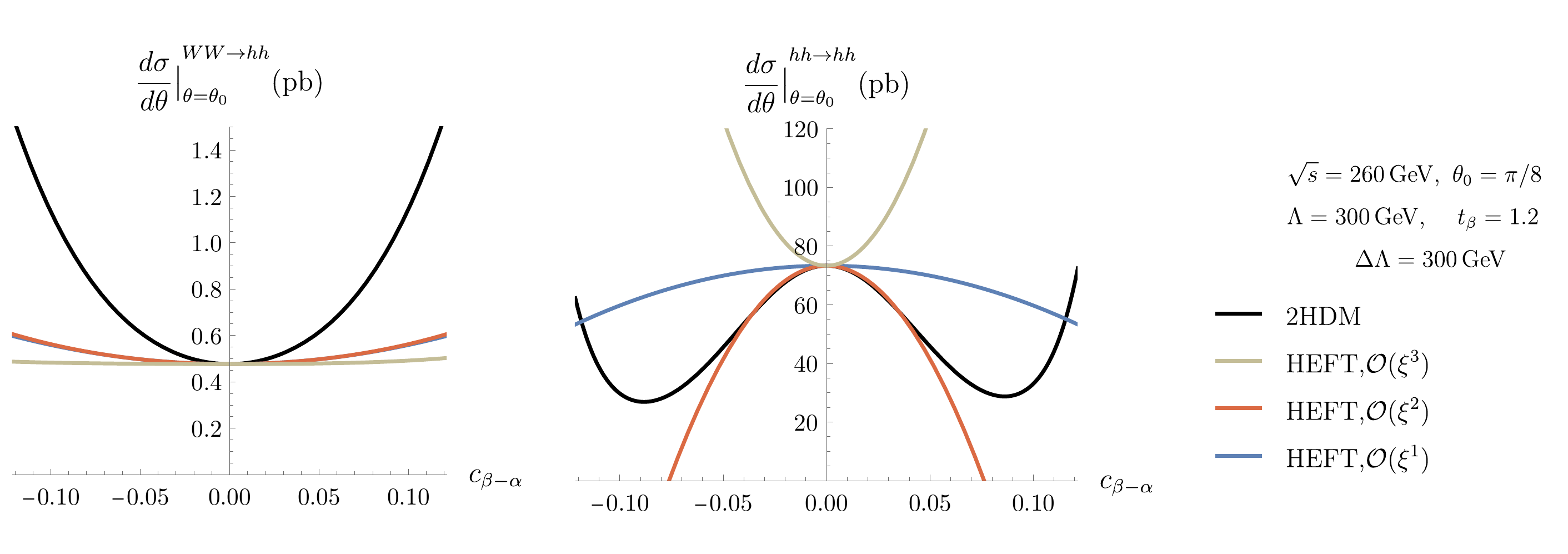}
\vspace{-10mm}
\caption{The same as in figure~\ref{fig:hhhh-HEFT-vs-2HDM}, but for $\Lambda=\Delta \Lambda=300$ GeV (on the left plot, the blue is behind the red). As before, the region of values of $c_{\beta-\alpha}$ shown is allowed by the theoretical constraints.}
\label{fig:hhhh-HEFT-vs-2HDM-yet}
\end{figure}

In figure~\ref{fig:hhhh-HEFT-vs-2HDM-yet}, we investigate the scenario in which the decoupling is lost. These plots are equivalent to those of figure~\ref{fig:hhhh-HEFT-vs-2HDM}, 
but with $\Lambda = \Delta \Lambda = 300$ GeV. Note that, even if this means $m_H = m_A = m_{H^+} = 600$ GeV, the choice $\Lambda=300$ GeV is a blatant violation of the assumptions of eq.~(\ref{eq:decoupling_def}). Indeed, both plots of figure~\ref{fig:hhhh-HEFT-vs-2HDM-yet} clearly show that the EFT is no longer valid according to the expansion of eqs.~(\ref{eq:decoupling_def}) and~(\ref{eq:scaling}): the different orders do not improve the convergence to the 2HDM results. We verified that the same conclusion holds for even smaller values of $\Lambda$.

%% file: conclusions.tex
\label{conclusions}

In this work, we presented two EFT matchings to the 2HDM: the SMEFT and the HEFT. 
We began with the 2HDM as our UV complete theory and imposed decoupling and perturbativity on the model.  This implies
that in the large mass limit of the heavy Higgs masses, the mixing angle $c_{\beta-\alpha}$ must obey the scaling $c_{\beta-\alpha}\sim \xi$,
where $\xi$ parameterizes the approach to the alignment limit, $c_{\beta-\alpha}\rightarrow 0$.
We  organized our studies of the SMEFT and HEFT matching in terms of an expansion in powers of $\xi$.

We discussed the matching of the HEFT to the 2HDM at $\mathcal{O}(\xi^2)$ (the matching of the SMEFT to that order was discussed in a previous paper,~\cite{Dawson:2022cmu}) and 
 used the unitary gauge to simplify the results, which were checked in an arbitrary $R_\xi$ gauge.
 The  matching equations for the parameters of the HEFT Lagrangian relevant for the processes discussed in this paper were given analytically.

We found that the SMEFT and the HEFT matchings to the 2HDM were identical to ${\cal{O}}(\xi^2)$ when the UV theory is required to obey decoupling and perturbativity. This holds for the fits to global Higgs signal strength, as well as the tree-level scatterings $WW\rightarrow hh$ and $hh\rightarrow hh$.
We investigated how accurately the  EFT matching at $\mathcal{O}(\xi^2)$ reproduces the 2HDM results in both these scatterings. In $WW\rightarrow hh$, the EFT matching accurately reproduces the 2HDM result, with differences smaller than the percent level. In the case of $hh\rightarrow hh$, by contrast, it fails to properly reproduce the 2HDM result in some regions of the parameter space. In this case, therefore, even the second order of the SMEFT (HEFT) expansion is not enough, and one should in principle consider dimension-10 operators (next-to-leading order operators in $p^2$). We further showed that
the convergence to the 2HDM could be improved if $\mathcal{O}(\xi^3)$ effects are included in the HEFT. Finally, we probed the case without decoupling, and concluded that the EFT expansion 
in powers of $\xi$ does not converge in that case.

This paper is a first exploration of the matchings of a UV model to both the SMEFT and the HEFT in a way consistent with decoupling and perturbativity. Several directions of future work are open. 
It would be particularly interesting to ascertain if the similarities between the two approaches found here will also hold for higher orders in the EFT expansion, as well for other processes. 
UV models other than the 2HDM could also be explored, with the purpose of ascertaining the consequences of pertubativity for the matchings in those cases. Also interesting would be the study of loops, and their impact in the comparison between the SMEFT and the HEFT matchings to a UV model~\cite{Buchalla:2022vjp}.

\textbf{Note added:} as this paper was being finished, ref.~\cite{banta:2023prj} was made publicly available. It focuses on the SMEFT matching to the 2HDM and proposes a basis alternative to the Higgs basis. That reference is an interesting complement to our paper.

%% file: appendix-2HDM.tex
\label{2hdm-appendix}

The quartic couplings of eq.~(\ref{eq:potential}) can be written in terms of the parameters of eq.~(\ref{eq:indep-real}) as:
\bs
\label{eq:theZs-real}
\bea
\label{eq:Z1}
Z_1 &=& \dfrac{\left(1 - c_{\beta-\alpha}^2\right) m_{h}^2 + c_{\beta-\alpha}^2 \, m_{H}^2}{v^2},
\label{eq:zmass1} \\
\label{eq:Z2}
Z_2 &=&
\dfrac{1}{2 \, v^2 \, t_{\beta}^3} \bigg[
c_{\beta-\alpha}^2 \, t_{\beta} \left(3 t_{\beta}^4 - 8 t_{\beta}^2 + 3\right)
   \left(m_{h}^2 - m_{H}^2\right) + \sqrt{1 - c_{\beta-\alpha}^2} \,
   c_{\beta-\alpha} \Big(t_{\beta}^6 - 7 t_{\beta}^4 + 7 t_{\beta}^2 \nonumber\\
   && \hspace{13mm} -1\Big) \left(m_{h}^2-m_{H}^2\right) - m_{h}^2
   \left(t_{\beta}^5 - 4 t_{\beta}^3 + t_{\beta}\right) + 2 t_{\beta}
   \left(t_{\beta}^2 - 1\right)^2
   \left(m_{H}^2 - Y_2\right)
\bigg],
\\
Z_3 &=& \dfrac{2}{v^2} \left(m_{H^{\pm}}^2 - Y_2\right), \\
Z_4 &=& \dfrac{c_{\beta-\alpha}^2 \left(m_{h}^2 - m_{H}^2\right) + m_A^2 + m_{H}^2 - 2 \, m_{H^{\pm}}^2}{v^2}, \\
Z_5 &=& \dfrac{c_{\beta-\alpha}^2 \left(m_{h}^2 - m_{H}^2\right) - m_A^2 + m_{H}^2}{v^2}, \\
Z_6 &=& \dfrac{c_{\beta-\alpha} \, \sqrt{1-c_{\beta-\alpha}^2} \, \left(m_{h}^2 - m_{H}^2\right)}{v^2}, \\
\label{eq:Z7}
Z_7 &=& \dfrac{1}{2 \, v^2 \, t_{\beta}^2} \Bigg[
-3 c_{\beta-\alpha}^2 t_{\beta} \left(t_{\beta}^2-1\right)
   \left(m_{h}^2-m_{H}^2\right)-\sqrt{1-c_{\beta-\alpha}^2}
   c_{\beta-\alpha} \left(t_{\beta}^4-4 t_{\beta}^2+1\right)
   \left(m_{h}^2-m_{H}^2\right) \nonumber \\
   && \hspace{13mm} +t_{\beta}
   \left(t_{\beta}^2-1\right) \left(m_{h}^2-2 m_{H}^2+2
   Y_2\right)
\Bigg].
\label{eq:zmass}
\eea
\es
As mentioned in section~\ref{2hdm}, the $\mathbb{Z}_2$ symmetry implies that not all the $Z$'s are independent. The two dependence relations read~\cite{Belusca-Maito:2016dqe}:
\bs
\bea
Z_2-Z_1 &=& \frac{1-2 s_\beta^2}{s_\beta c_\beta}\left(Z_6+Z_7\right), \\
Z_{345}-Z_1 &=& \frac{1-2 s_\beta^2}{s_\beta c_\beta} Z_6-\frac{2 s_\beta c_\beta}{1-2 s_\beta^2}\left(Z_6-Z_7\right).
\eea
\es
Eqs.~(\ref{eq:2HDM-masses-sq-a}) and~(\ref{eq:2HDM-masses-sq-b}) can be rewritten by replacing the dependence on $c_{\beta-\alpha}$ by $Z_i$ parameters as:
\bs
\label{eq:2HDM-masses-sq-var}
\bea
\label{eq:2HDM-masses-sq-var-a}
m_h^2 &=& \dfrac{2 \, Y_2 + v^2 (2 Z_1 + Z_{345}) - 
    \sqrt{\big[2 \, Y_2 + v^2 (Z_{345} - 2 Z_1)\big]^2 + 16 v^4 Z_6^2}}{4}, \\
\label{eq:2HDM-masses-sq-var-b}
m_H^2 &=& \dfrac{2 \, Y_2 + v^2 (2 Z_1 + Z_{345}) + 
    \sqrt{\big[2 \, Y_2 + v^2 (Z_{345} - 2 Z_1)\big]^2 + 16 v^4 Z_6^2}}{4}.
\eea
\es

In what follows, we present futher details concerning the integration out of $H$, $A$ and $H^+$ in the HEFT. 
As we saw in section~\ref{heft}, the use of the unitary gauge implies a maximum of four fields in each term of the 2HDM Lagrangian. 
Then, from eq.~(\ref{eq:L2hdmjs}), it is clear that $J_0$, $J_1$, $J_2$, $J_3$, and $J_4$ will contain only light fields with a maximum number of four, three, two, one, and zero, respectively.
For the tree-level scattering processes $WW\to hh$ and $hh\to hh$, only the following $J$'s are need: $J_0$ (up to four light fields) and $J_1^a$, with $a=H,H^+$ (up to two light fields).
They read:
\begin{eqnarray}
\label{eq:j0}
J_0 &=&
\frac{1}{2} \partial _{\mu }(h){}^2-\frac{1}{2} h^2 m_h^2 
  + \left(\frac{1}{2}m_Z^2 Z^{\mu }Z_\mu +m_W^2 W^{\mu } W_{\mu }^{\dagger }\right) \,\left(1 + \frac{2 s_{\beta-\alpha} h}{v} +\frac{h^2}{v^2}\right)  
\nonumber\\  
&&+\frac{1}{4 v t_{\beta }^2} h^3 \bigg\{\left(t_{\beta }^4-4 t_{\beta }^2+1\right) c_{\beta-\alpha }^4 \left(m_h^2-m_H^2\right) s_{\beta- \alpha}+3 t_{\beta } \left(t_{\beta }^2-1\right) c_{\beta-\alpha }^5
   \left(  m_h^2-m_H^2  \right)\nonumber\\
   && -t_{\beta } \left(t_{\beta }^2-1\right) c_{\beta-\alpha }^3 \left(m_h^2-2 m_H^2+2 Y_2\right) - 2 t_{\beta }^2 c_{\beta-\alpha }^2 \left(m_h^2-2 Y_2\right)
   s_{\beta -\alpha } - 2 m_h^2 t_{\beta }^2 s_{\beta -\alpha }\bigg\}
\nonumber\\
&& -\frac{1}{16 v^2 t_{\beta }^3}h^4 \bigg\{\big[t_{\beta }^6-19 t_{\beta }^4+19 t_{\beta }^2-1\big] c_{\beta-\alpha
   }^5 \left(m_H^2-m_h^2\right) s_{\beta -\alpha }\nonumber\\
   && -4 t_{\beta }^2 \left(t_{\beta }^2\right) c_{\beta-\alpha }^3 s_{\beta -\alpha} \left(m_h^2-2 m_H^2+2 Y_2\right)+t_{\beta } \left(7
   t_{\beta }^4-26 t_{\beta }^2+7\right) c_{\beta-\alpha }^6 \left(m_h^2-m_H^2\right)\nonumber\\
   && +t_{\beta } c_{\beta-\alpha }^4 \big[m_h^2 \left(-5 t_{\beta }^4+18 t_{\beta }^2-5\right)+6 m_H^2
   \left(t_{\beta }^4-4 t_{\beta }^2+1\right)-2 Y_2 \left(t_{\beta }^4-6 t_{\beta }^2+1\right)\big]\nonumber\\
   && +2 t_{\beta }^3 c_{\beta-\alpha }^2 \left(m_h^2+m_H^2-4 Y_2\right)+2 m_h^2 t_{\beta
   }^3\bigg\} \, , \\
\label{eq:j1k2H}
J_1^{H} &=&   
 \frac{2 c_{\beta-\alpha }}{ v}  \left(  m_W^2 W^{\mu } W_{\mu }^{\dagger }+\frac{1}{2}m_Z^2 Z^{\mu }Z_\mu\right) \nonumber    
\\
&&   + \frac{c_{\beta-\alpha }}{4 v t_{\beta }^2} h^2 \bigg\{        
   9 t_{\beta } \left(t_{\beta }^2-1\right) c_{\beta-\alpha }^3 \left(m_h^2-m_H^2\right) s_{\beta -\alpha } 
\nonumber\\
&& -3 t_{\beta } \left(t_{\beta
   }^2-1\right) c_{\beta-\alpha } s_{\beta-\alpha } \left(m_h^2-2 m_H^2+2 Y_2\right)+3 \left(t_{\beta }^4-4 t_{\beta }^2+1\right) c_{\beta-\alpha }^4
   \left(m_h^2-m_H^2\right)\nonumber\\
   && +c_{\beta-\alpha }^2 \big[m_h^2 \left(-3 t_{\beta }^4+8 t_{\beta }^2-3\right)+m_H^2 \left(3 t_{\beta }^4-14 t_{\beta }^2+3\right)+12 Y_2 t_{\beta
   }^2\big]+2 t_{\beta }^2 \left(m_H^2-4 Y_2\right)
    \bigg\} \nonumber\\
&& + \mO(h^3)\, , \\
\label{eq:j1k2HP}
J_1^{H^{+}} &=& \left(J_1^{H^-}\right)^\dagger =   
\frac{i m_W c_{\beta -\alpha }}{v} \bigg[h \left( \partial^{\mu }W_{\mu }\right)+2 W^{\mu } \left( \partial _{\mu }h \right)\bigg] \, ,
\end{eqnarray} 
where we express $e$, $c_W$ and $s_W$ by means of $g=e/s_W=2 m_W/v$ and $c_W=m_W/m_Z$. 
The case $a=A$, with $J_1^{A} = -c_{\beta -\alpha } \left[h \left(\partial ^{\mu }Z_{\mu }\right)+2 Z^{\mu } \left(\partial _{\mu }h \right)\right] \left(m_W^2+m_Z^2 s_W^2\right)/(v m_Z)$ would contribute to the $ZZ\to hh$ process that is  not studied here.

%% file: appendix-singlet.tex
\label{singlet-appendix}

We briefly review the $\mathbb{Z}_2$ symmetric real singlet extension of the SM discussed in ref.~\cite{Buchalla:2016bse} in the context of the HEFT matching. Our purpose is to illustrate the crucial differences between that model and the 2HDM. The potential in terms of a real singlet, $S$, and the usual $SU_L(2)$ doublet, $\phi$, is
\begin{eqnarray}
    V&=&
    -{\mu_1^2\over 2} \phi^\dagger\phi -{\mu_2^2\over 2} S^2
    +{\lambda_1\over 4}(\phi^\dagger \phi)^2
    +{\lambda_2\over 4} S^4 +{\lambda_3\over 2} \phi^\dagger \phi \, S^2
    \, .
\end{eqnarray}
After SSB, $\phi$ and $S$ get vevs $v/\sqrt{2}$ and $v_s/\sqrt{2}$, respectively.
The physical states $h$ and $H$ have masses $m$ and $M$, respectively ($m$ is assumed to be light and $M$ heavy). These are determined by minimizing the potential and diagonalizing the mass matrix with the mixing angle $\chi$.
This happens such that the Feynman rule for the cubic self-interaction of $h$ reads:
\be
i \dfrac{m^2}{2 v v_s} (s_{\chi}^3 \, v - c_{\chi}^3 \, v_s).
\ee
Therefore, in stark contrast with what happens in the 2HDM (recall eq.~(\ref{eq:triple-Higgs})), the cubic self-interaction of $h$ in the model of ref.~\cite{Buchalla:2016bse} does not scale with positive powers of heavy masses (in this case, just $M$).
This allows the authors to perform a HEFT matching as an expansion in inverse powers of the heavy mass $M$.
On the other hand, such an expansion does not comply with decoupling and perturbativity.%
\fn{This does not mean that the different orders in the $1/M$ expansion performed in ref.~\cite{Buchalla:2016bse} violate perturbativity. The problem, rather, is that the expansion \textit{itself} does not respect perturbativity for a very large $M$, if no other assumption is made.}
To see this, note that the quartic couplings of the potential can be written in terms of the masses, the vevs and the mixing angle as:
\be
\label{eq:lambdas}
\lambda_1 = \dfrac{2}{v^2} \Big[M^2 s_{\chi}^2 - m^2 (s_{\chi}^2-1) \Big],
\quad
\lambda_2 = \dfrac{2}{v_s^2} \Big[m^2 s_{\chi}^2 - M^2 (s_{\chi}^2-1) \Big],
\quad
\lambda_3 = \frac{2 \, c_{\chi} \, s_{\chi}}{v \, v_s} (M^2 - m^2).
\ee
This clearly shows that, if $M$ is taken to be very large and no other assumption is made, perturbativity is violated. As a consequence, even if no inconsistency is found in the cubic self-interaction of $h$, an expansion that simply assumes $M$ to be very large and uses $1/M$ as an expansion parameter does not comply with perturbativity.
Such compliance thus requires a different expansion, with more assumptions --- specifically, assumptions about $v_s$ and $\chi$. Along the lines of eq.~(\ref{eq:scaling}), the scalings $1/M^2 \sim \mathcal{O(\xi)}$, $1/v_s^2 \sim \mathcal{O(\xi)}$ and $s_{\chi}^2 \sim \mathcal{O(\xi)}$ would lead to well-behaved quartic couplings.